\documentclass[aps,prl,reprint,groupedaddress,superscriptaddress,floatfix]{revtex4-2}

\usepackage[T1]{fontenc}
\usepackage{newtxtext}
\usepackage{newtxmath}
\usepackage{bm}

\usepackage{titlesec}

\usepackage{dsfont}
\usepackage{mathtools}
\usepackage{physics}

\usepackage{xcolor}
\usepackage{soul}

\usepackage{graphicx}
\usepackage[colorlinks,citecolor=blue,linkcolor=blue,urlcolor=blue]{hyperref}
\usepackage{hypcap}

\usepackage{enumitem}

\usepackage{tabularx}
\usepackage{makecell}

\graphicspath{{img/}}

\usepackage{cleveref}
\usepackage{etoolbox}

\makeatletter
\let\addcontentslineOriginal\addcontentsline
\let\addcontentsline\@gobblethree
\makeatother

\begin{document}
% -------------------------------
% MANUSCRIPT
% -------------------------------
\title{Strong Eigenstate Thermalization from Mean-Ergodic Non-chaotic Dynamics}

\author{Avadhut V. Purohit}
\email{avdhoot.purohit@gmail.com}
\affiliation{Department of Physics, Visvesvaraya National Institute of Technology, Nagpur 440010, India}
\author{Harshit Sharma}
\email{harshitsharma2796@gmail.com}
\affiliation{Department of Physics, Visvesvaraya National Institute of Technology, Nagpur 440010, India}
\author{Udaysinh T. Bhosale}
\email{udaysinhbhosale@phy.vnit.ac.in}
\affiliation{Department of Physics, Visvesvaraya National Institute of Technology, Nagpur 440010, India}
\date{\today}

% \begin{abstract}
%     We report an example of a many-body system, derived from the double kicked top (DKT), with non-chaotic yet mean-ergodic dynamics that displays \textit{strong} eigenstate thermalization hypothesis (ETH) in the quantum regime. The analysis addresses a key open question: whether \textit{strong} ETH is a quantum analog of ergodicity (or mean-ergodicity). Despite non-chaotic dynamics, the fluctuations of the diagonal matrix elements of an observable scale as $D^{-1/2}$, where $D$ denotes the Hilbert space dimension. Furthermore, the off-diagonal matrix elements show Gaussian statistics together with a smooth function $f_O(\bar{E}, \omega)$ that becomes nearly uniform in the large-$k_\theta$ domain. Our findings show that even mean-ergodic and non-chaotic systems can exhibit \textit{strong} ETH.
% \end{abstract}
\begin{abstract}
    We report an example of a many-body system, derived from the double kicked top (DKT), with non-chaotic yet mean-ergodic dynamics that displays \textit{strong} eigenstate thermalization hypothesis (ETH) in the quantum regime. The analysis addresses a key open question: whether \textit{strong} ETH is a quantum analog of ergodicity (or mean-ergodicity). Despite non-chaotic dynamics, the fluctuations of the diagonal matrix elements of an observable scale as $D^{-1/2}$, where $D$ denotes the Hilbert space dimension. Furthermore, the off-diagonal matrix elements show parameter-independent distribution, together with a smooth function $f_O(\bar{E}, \omega)$ that becomes nearly uniform in the large-$k_\theta$ domain. Our findings show that even mean-ergodic and non-chaotic systems can exhibit \textit{strong} ETH.
\end{abstract}

\maketitle

\textit{Introduction --- }ETH provides a quantum explanation for the emergence of statistical mechanics~\cite{ruelle1969statistical,ornstein1989ergodic,deutsch1991quantum,srednicki1994chaos,zelditch2005quantum,rigol2008thermalization,reimann2015eigenstate,d2016quantum,deutsch2018eigenstate,yoshizawa2018numerical,pappalardi2022eigenstate,murthy2023non,SilviaThermalization2026}. It implies that individual eigenstates of such a quantum system are thermal. Such systems typically show Wigner-Dyson level statistics, while some deviations were observed due to integrability, additional conservation laws, or rare non-thermal eigenstates~\cite{srednicki1994chaos,rigol2008thermalization,d2016quantum,yoshizawa2018numerical}. Taken together, these observations have led to the widespread expectation that chaotic systems thermalize, whereas non-chaotic systems usually do not.

We would like to mention a few noteworthy exceptions to ETH. Some nonintegrable systems show failure of ETH due to effects such as many-body localization and quantum scars~\cite{heller1984bound,turner2018weak,bernien2017probing,robinson2019signatures,murthy2023non,srdinvsek2024ergodicity}. On the other hand, some integrable systems, such as the XXX Heisenberg model and generic translation-invariant systems, have been shown to obey a weaker form of ETH~\cite{alba2015eigenstate,pilatowsky2025quantum}. Studies have also shown that even weak integrability-breaking perturbations can lead to thermal behavior in dynamics~\cite{Brenes2020}. 

To better understand the relationship between dynamical thermalization and eigenstate properties, useful notions of weak and strong ETH were introduced~\cite{kim2014testing,Ishii2019Strong}. In \textit{weak} ETH, expectation values of an observable in \textit{almost all} eigenstates agree with the microcanonical average, whereas in \textit{strong} ETH, expectation values of an observable in \textit{all} eigenstates match the microcanonical average~\cite{biroli2010effect,kim2014testing,d2016quantum,Ishii2019Strong}.

It was shown that a few nonintegrable models display \textit{strong} ETH~\cite{kim2014testing}. In noninteracting integrable systems also, the \textit{strong} ETH behavior was observed in appropriately constrained subspaces~\cite{Ishii2019Strong}. Nevertheless, a broader question of whether \textit{strong} ETH can emerge from classical non-chaotic, ergodic dynamics remains open. This question is particularly relevant for systems with a well-defined semiclassical limit, where the connection between classical and quantum dynamics can be established more directly~\cite{haake1987classical,neill2016ergodic,ruebeck2017entanglement,purohit2025double}. Consequently, a non-chaotic but ergodic (or mean-ergodic~\cite{neumann1932proof,cornfeld2012ergodic,petersen1995ergodic}) system provides a natural setting to test whether \textit{strong} ETH can arise without chaos.

The DKT model, originally introduced in Refs.~\cite{haake1987classical,haake1991quantum} to study broken time-reversal symmetry in quantum chaos, offers a natural platform for such dynamics~\cite{purohit2025double}. In this model, an additional periodic nonlinear kick is applied perpendicular to the linear precession and the first nonlinear kick. The dynamics of the DKT can be better understood in terms of two transformed kick strengths $(k, k') \rightarrow (k_r, k_\theta)$~\cite{purohit2025double}. Here, the parameter $k_r$ was observed to be equivalent to the kicking strength $k$ of the standard kicked top, while $k_\theta$ breaks time-reversal symmetry. The study showed that the parameter $k_\theta$ induces no bifurcations, yet increases the long-time-averaged entanglement entropy~\cite{purohit2025double}. These features motivate us to analyze classical (non-chaotic but mean-ergodic) and quantum dynamics (ETH in particular) as a function of $k_\theta$.

In this Letter, we analyze the DKT in the near-integrable regime $k_r \leq 1$, particularly at $k_r = 1$. We find that $k_\theta$ does not alter the degree of chaos but induces mean-ergodic motion for large values in the domain $k_\theta \in (0, 10^6)$. In the corresponding quantum regime, we show \textit{strong} ETH behavior. Although we demonstrate these findings only in the DKT model, the underlying dynamics (both classical and quantum) are not model-specific. Therefore, our analysis suggests that the mean-ergodicity is sufficient for the \textit{strong} ETH behavior.

\textit{Chaos independence of $k_\theta$ --- }The Floquet operator for the DKT model is given by~\cite{haake1991quantum}
\begin{equation}\label{eq:floquet}
    \mathcal{U} = \exp\!\left(- i \frac{k'}{2j}J_x^2\right)\exp\!\left(- i \frac{k}{2j}J_z^2\right)\exp\!\left(- i \frac{\pi}{2} J_y\right).
\end{equation}
The classical map $\mathbf{X}'=\mathbf{F}(\mathbf{X})$ can be obtained by using $\mathbf{J}' = \mathcal{U}^\dagger \mathbf{J} \mathcal{U}$ and taking $\lim_{j\to\infty}\mathbf{J}/j = \mathbf{X}$~\cite{purohit2025double} as follows:
\begin{widetext}
  \begin{align}\label{eq:classicalMap}
    X' &= Z \cos(k X) + Y \sin(k X),\\
    Y' &= \left[Y \cos(k X) - Z \sin(k X)\right] \cos\!\left[k'\!\left[Z \cos(k X) + Y \sin(k X)\right]\right] + X \sin\!\left[k'\!\left[Z \cos(k X) + Y \sin(k X)\right]\right],\\
    Z' &= - X \cos\!\left[k'\!\left[Z \cos(k X) + Y \sin(k X)\right]\right] + \!\left[Y \cos(k X) - Z \sin(k X)\right] \sin\!\left[k'\!\left[Z \cos(k X) + Y \sin(k X)\right]\right].
  \end{align}
\end{widetext}
The transformation given by
\begin{align}
    k_{r} = \frac{k + k'}{2}, \quad k_{\theta} = \frac{k - k'}{2},
\end{align}
simplifies the dynamics into two parts: QKT equivalent dynamics given by $k_r$ and the broken time-reversal symmetric part given by $k_\theta$. The first set of non-trivial fixed points is obtained by solving $\mathbf{F(X)}=\mathbf{X}$ as follows~\cite{purohit2025double}:
\begin{align}\label{eq:fixedpoint}
    Z &= -X \sin\left(k_\theta X\right)\csc\left(k_r X\right),\notag \\
    Y &= X \cos\left(k_\theta X\right)\csc\left(k_r X\right) \, \text{and} \\
    f(X) &= \frac{\sin^2\left(k_r X\right)}{1 + \sin^2\left(k_r X\right)} - X^2 = 0.\notag
\end{align}
Notice that the parameter $k_\theta$ only controls the location of a fixed point through $Z$ and $Y$. 
Whereas $k_r$ governs the occurrence of fixed points by numerically solving the transcendental equation $f(X)=0$. This suggests that the parameter $k_\theta$ does not play any role in local bifurcation analysis (see Sec.~II of Ref.~\cite{purohit2025double} for dynamics). It further implies that if at all there is any $k_\theta$-dependence of chaos, it would arise only through global mechanisms.

For a given $k_r$, the parameter $k_\theta$ rotates and stretches phase-space structures around the trivial fixed points (see Figs.~(2), (4), (6) and (7) of Ref.~\cite{purohit2025double}). This deformation depends on the distance from the trivial fixed points. Points close to $X=\pm 1$ move slowly, while farther points move faster with $k_\theta$. As a result, phase-space structures experience stretching as $k_\theta$ increases (see Fig.~\ref{fig:phase_space}, Fig.~\ref{supp:fig:phase_space} of supplementary material, and Figs.~(2), (4), (6) and (7) of Ref.~\cite{purohit2025double}). For sufficiently large $k_\theta$, this stretching causes trajectories to spread across the accessible (limit $k_{\theta}\rightarrow \infty$) phase-space without disturbing the stability, as illustrated in Fig.~\ref{fig:phase_space}~(b). 
\begin{figure}
    \includegraphics[width=\linewidth]{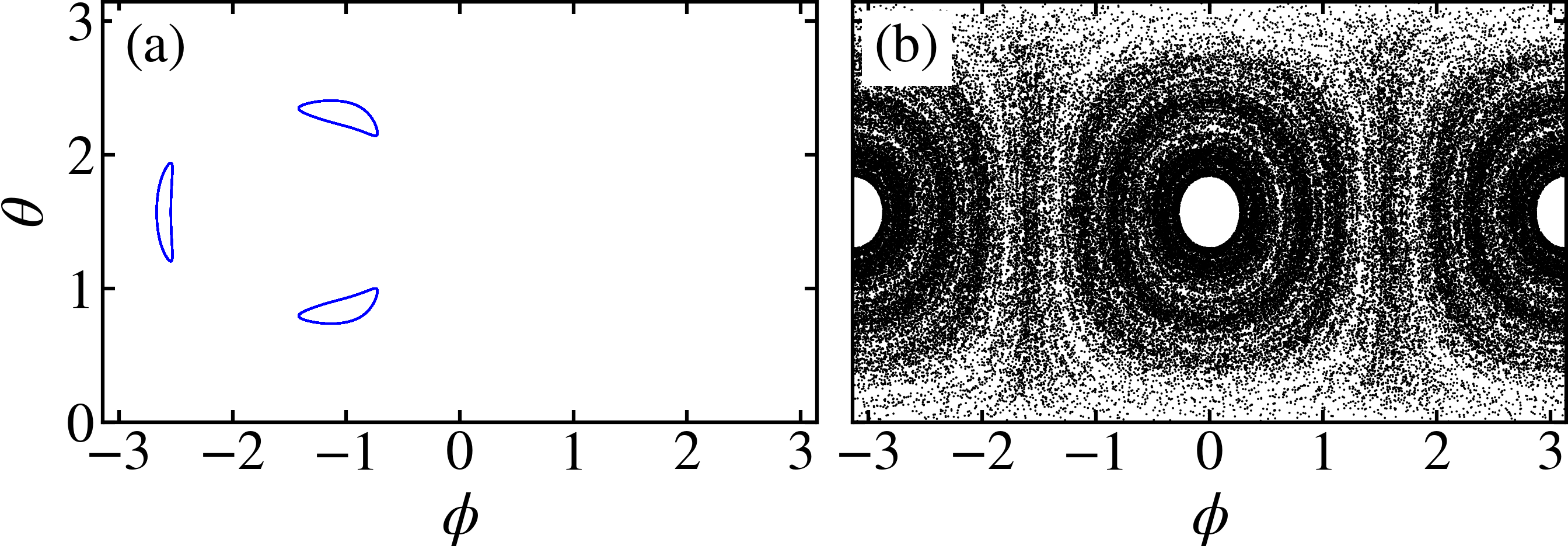}
    \caption{The phase-space portrait for the initial point $(\theta_0 = 1.0, \phi_0 = -0.75)$ evolved for $10^5$ kicks. Here, $k_r = 1$, (a) $k_\theta = 0$ and (b) $k_\theta = 10^4$. Blue dots indicate a period-3 cycle in the left panel. The right panel shows non-chaotic, mean-ergodic evolution of the same initial point illustrated by black dots.}\label{fig:phase_space}
\end{figure}

\begin{figure}
    \includegraphics[width=0.48\linewidth]{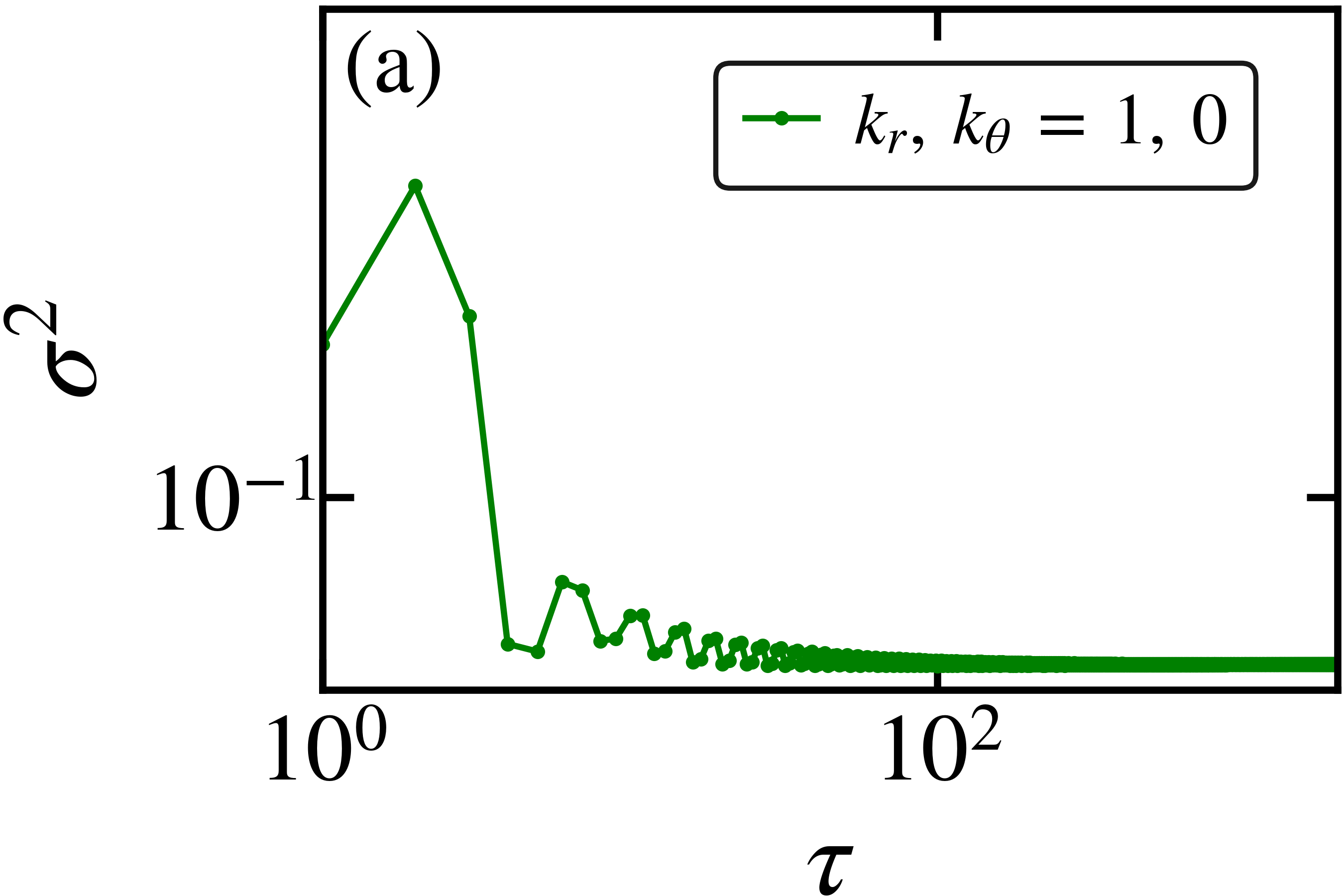}
    \includegraphics[width=0.48\linewidth]{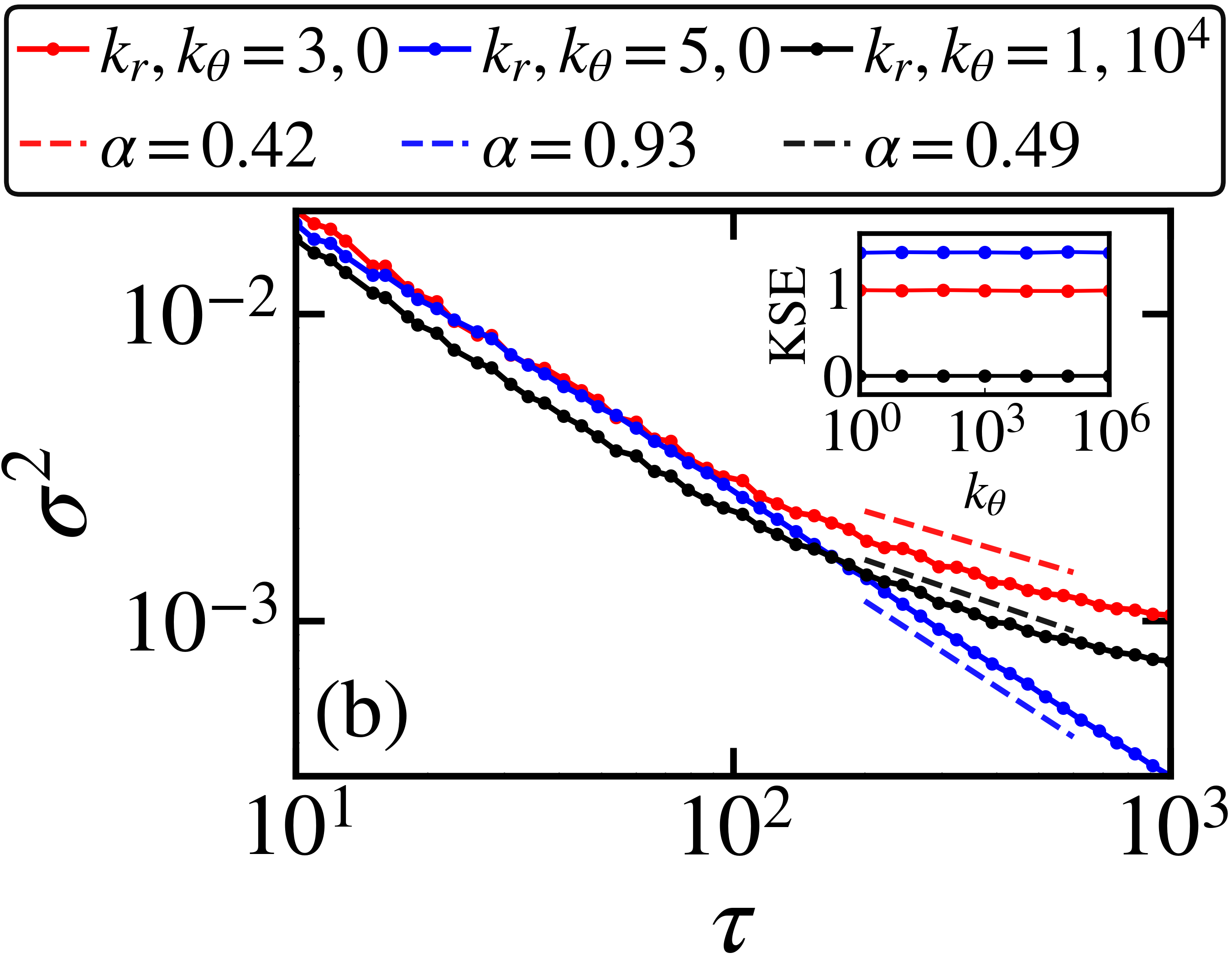}
    \caption{The variance of time-averaged trajectory lengths $\operatorname{Var}_\tau(l)$ plotted versus $\tau$ using $M = 100 \times 100$ equally spaced initial conditions from the phase-space. Panel (a) shows the non-mean-ergodic case $(k_r = 1, k_\theta = 0)$, where variance saturate to a non-zero value. Panel (b) shows mean-ergodic cases fitted with the power-law $\operatorname{Var}_\tau(l) \propto \tau^{-\alpha}$ for two chaotic cases: red $(k_r = 3, k_\theta = 0)$ and blue $(k_r = 5, k_\theta = 0)$, together with the non-chaotic, mean-ergodic case: black $(k_r = 1, k_\theta = 10^4)$. The inset shows the KSE versus $k_\theta$ for the initial condition $(\pi/2, - \pi/2 - 0.1)$. Black, red, and blue dots correspond to $k_r = 1$, $k_r = 3$, and $k_r = 5$, respectively.}\label{fig:mean-ergodicity}
\end{figure}

Although $k_\theta$ was previously associated with weak changes in chaos through the Kolmogorov-Sinai entropy (KSE)~\cite{purohit2025double}, our extended computations show no detectable dependence of the KSE on $k_\theta$ even up to $k_\theta = 10^6$ for the initial condition $(\pi/2,-\pi/2-0.1)$ [see inset of Fig.~\ref{fig:mean-ergodicity}(b)]. This observation is consistent with the analytical expansion of Eq.~(32) in Ref.~\cite{purohit2025double}, which predicts no leading-order contribution from $k_\theta$. These results support the conclusion that the degree of chaos remains effectively independent of $k_\theta$ within the studied domain.

\textit{Mean-ergodicity --- }\label{sub:sec:ergodicity}
Having demonstrated $k_\theta$ as a parameter that does not change the order of chaos in the studied domain, we now investigate its role in ergodic convergence. For large-$k_\theta$, long-time evolution leads to significant spreading of trajectories across the accessible region of phase-space (see Fig.~\ref{fig:phase_space}~(b)). However, such spreading alone does not guarantee pointwise (Birkhoff) ergodicity~\cite{petersen1995ergodic,cornfeld2012ergodic}, since time-averaged observables may still depend on the initial conditions (see Fig.~\ref{supp:fig:phase_space} in supplementary material). Accordingly, we analyze convergence through ensemble fluctuations of time-averaged observables~\cite{rebenshtok2008weakly,carmi2011fractional,cornfeld2012ergodic,barkai2012strange}. Here, mean-ergodic convergence (von Neumann's quasi-ergodicity) refers to convergence of time-averaged observables in the \textit{mean-square} sense, rather than \textit{almost-everywhere} convergence in the pointwise sense (see related discussion on page 34 of Ref.~\cite{cornfeld2012ergodic} and Sec. 2.1 of Ref.~\cite{petersen1995ergodic}). Operationally, this corresponds to the decay of ensemble fluctuations of time-averaged observables toward zero, even when individual trajectories do not necessarily exhibit pointwise convergence.

Since the trajectory length on the compact phase-space is continuous and defined for all initial conditions, we verify mean-ergodic convergence by analyzing the ensemble fluctuations of time-averaged trajectory lengths. We have also verified it for other observables such as the $y$-coordinate (see Fig.~\ref{supp:fig:ergodicity} in the supplementary material). The time-averaged trajectory length and its ensemble average are given as follows:
\begin{align}
    \bar{l}_\tau = \frac{1}{\tau} \sum_{n=0}^{\tau -1} \left|\mathbf{X}_n - \mathbf{X}_0\right| \; \text{ and }\;
    {\langle l_\tau\rangle}_\text{ens} = \frac{1}{M} \sum_{i=1}^{M} \bar{l}^{(i)}_{\tau},
\end{align}
where $M$ is the number of equispaced initial conditions considered. The variance is given as follows (see discussion above Eq.~(48) in Ref.~\cite{carmi2011fractional}, and Sec.~\ref{supp:sub:sec:analogy} of supplementary material):
\begin{align}\label{eq:variance}
    \operatorname{Var}_\tau(l) = {\langle l_\tau^2 \rangle}_\text{ens} - {\langle l_\tau \rangle}_\text{ens}^2.
\end{align}
In case of mean-ergodic convergence, $\operatorname{Var}_\tau(l) \to 0$ as $\tau \to \infty$, indicating that the variance of time-averaged trajectory lengths across initial conditions vanishes asymptotically. In systems with anomalous ergodic convergence, the variance often decays as $\tau^{-\alpha}$ with $0 < \alpha \leq 1$~\cite{rebenshtok2008weakly,carmi2011fractional,barkai2012strange}. For non-ergodic dynamics, the variance converges to a non-zero value as $\tau \to \infty$.

Our computations show that for the case $(k_r = 1, k_\theta = 0)$, the variance quickly approaches a non-zero constant value, indicating non-ergodic behavior (see Fig.~\ref{fig:mean-ergodicity}~(a)). In contrast, all other cases show decay of variance towards zero, implying mean-ergodic convergence of time-averaged observables. Thus, the large-$k_\theta$ domain at $k_r = 1$ corresponds to a non-chaotic but mean-ergodic phase.

The strongly chaotic case $(k_r = 5, k_\theta = 0)$ approaches the conventional $1/\tau$ scaling with $\alpha \approx 1$, indicating rapid decay of correlations. The weakly chaotic case $(k_r = 3, k_\theta = 0)$ shows slower convergence with $0 < \alpha < 1$. Interestingly, the non-chaotic case $(k_r = 1, k_\theta = 10^4)$ also exhibits a similar anomalous decay despite the absence of chaos. We emphasize that the exponent $\alpha$ characterizes the rate of mean-ergodic convergence rather than the presence or absence of chaos.

The slow convergence in the large-$k_\theta$ domain originates from the stretching and reorientation of phase-space structures without destabilizing them. Consequently, trajectories continue to explore correlated regions of phase-space over long times, leading to persistent memory effects and anomalously slow convergence of time-averaged observables. Since the observable $l_\tau$ explicitly depends on the initial point $\mathbf{X}_0$, its convergence is particularly sensitive to the rate at which correlations decay. Using observables that do not explicitly retain information about $\mathbf{X}_0$, we recover $\alpha \approx 1$ for all mean-ergodic cases (see Fig.~\ref{supp:fig:ergodicity} in the supplementary material). Therefore, although the chaotic and non-chaotic, mean-ergodic regimes show similar anomalously~\cite{metzler2014anomalous} slow convergence exponents for the observable considered, the underlying mechanisms are fundamentally different: the former originates from rapid chaos-induced decorrelation, while the latter emerges from long-time correlated spreading induced by large $k_\theta$. These results demonstrate mean-ergodic convergence of time-averaged observables in the non-chaotic large-$k_\theta$ domain.

\textit{ETH --- }The early signs of ETH for the case $(k_r = 1, \text{large-}k_\theta)$ can be observed in the earlier study of DKT (see Fig. (25) of~\cite{purohit2025double}), where long-time-averaged von Neumann entropy for a single-qubit reduced density matrix was analyzed. The study showed saturation of the entanglement to the maximum for large-$k_\theta$ in the large system-size $2j$ limit (see Fig.~\ref{supp:fig:vn_entropy}~(b) in the supplementary material), giving evidence of thermalization~\cite{kaufman2016quantum}.

We now demonstrate the \textit{strong} ETH by analyzing the matrix elements of an observable~$\hat{O}$ in the eigenstate $\lbrace | \phi_\alpha \rangle \rbrace$ of the Floquet operator $\mathcal{U}$. The ETH ansatz is given by
\begin{align}
    O_{\alpha\beta} = O\!\left(\bar{E}\right) \delta_{\alpha\beta} + e^{-S(\bar{E})/2} f_O\!\left(\bar{E}, \omega\right) R_{\alpha\beta},
\end{align}
where $\bar{E} = (E_\alpha + E_\beta)/2$, $\omega = E_\alpha - E_\beta$, $S(\bar{E})$ is the thermodynamic entropy at energy $\bar{E}$, $O(\bar{E})$ and $f_O(\bar{E}, \omega)$ are smooth functions of their arguments, and $R_{\alpha\beta}$ are random variables with zero mean and unit variance~\cite{d2016quantum,leblond2019entanglement,wang2024eigenstate,singha2025unveiling}. Due to the permutation symmetry of the DKT, we focus on the observable $O = J_z^2/j(j+1)$, which captures average two-body correlations over all pairs. The diagonal elements $O_{\alpha\alpha}$ represent the expectation value of the observable in the eigenstate $|\phi_\alpha\rangle$, while the off-diagonal elements $O_{\alpha\beta}$ with $\alpha \neq \beta$ determine the temporal fluctuations of the observable in the long time.

\begin{figure}
    \includegraphics[width=0.49\linewidth]{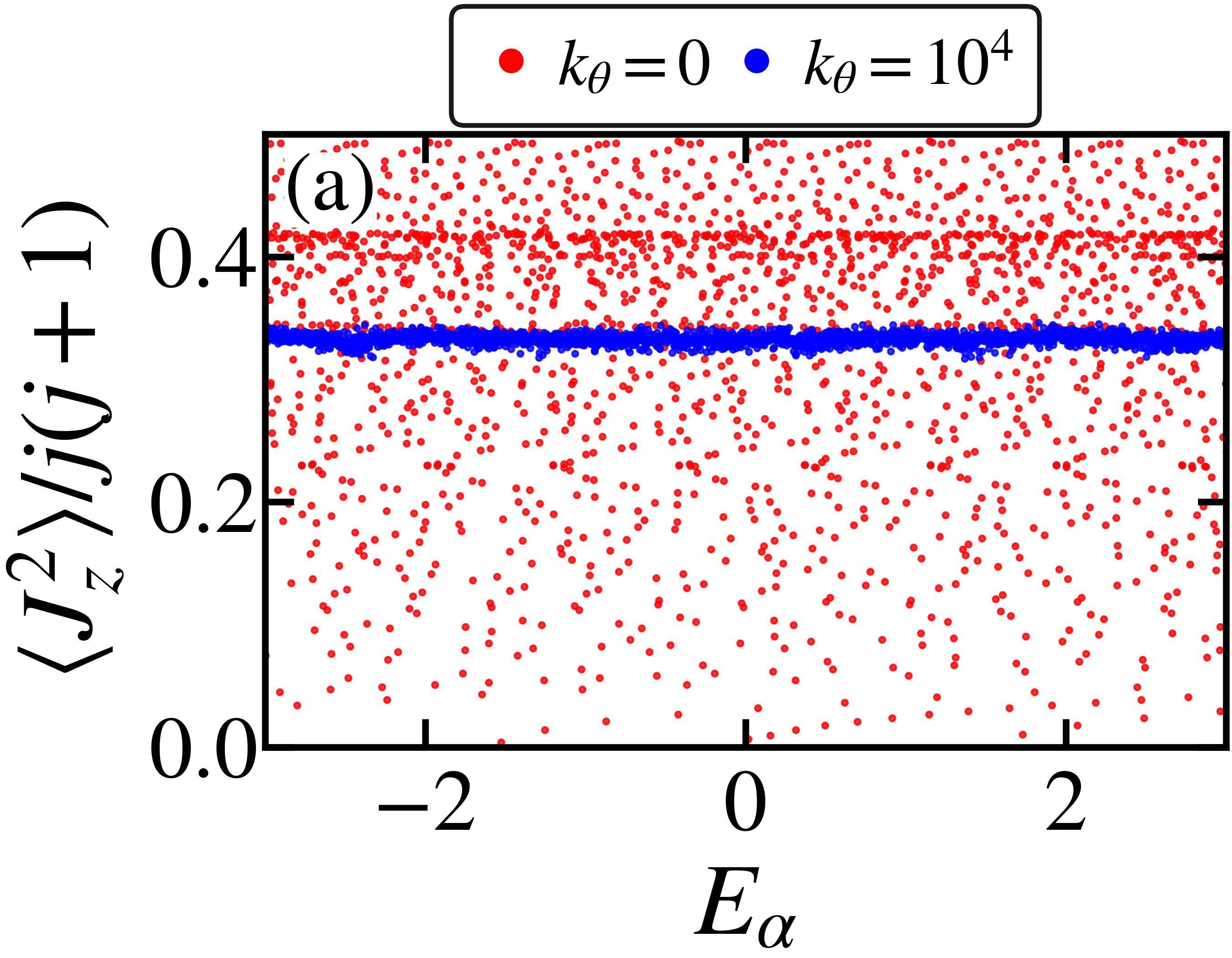}
    \includegraphics[width=0.49\linewidth]{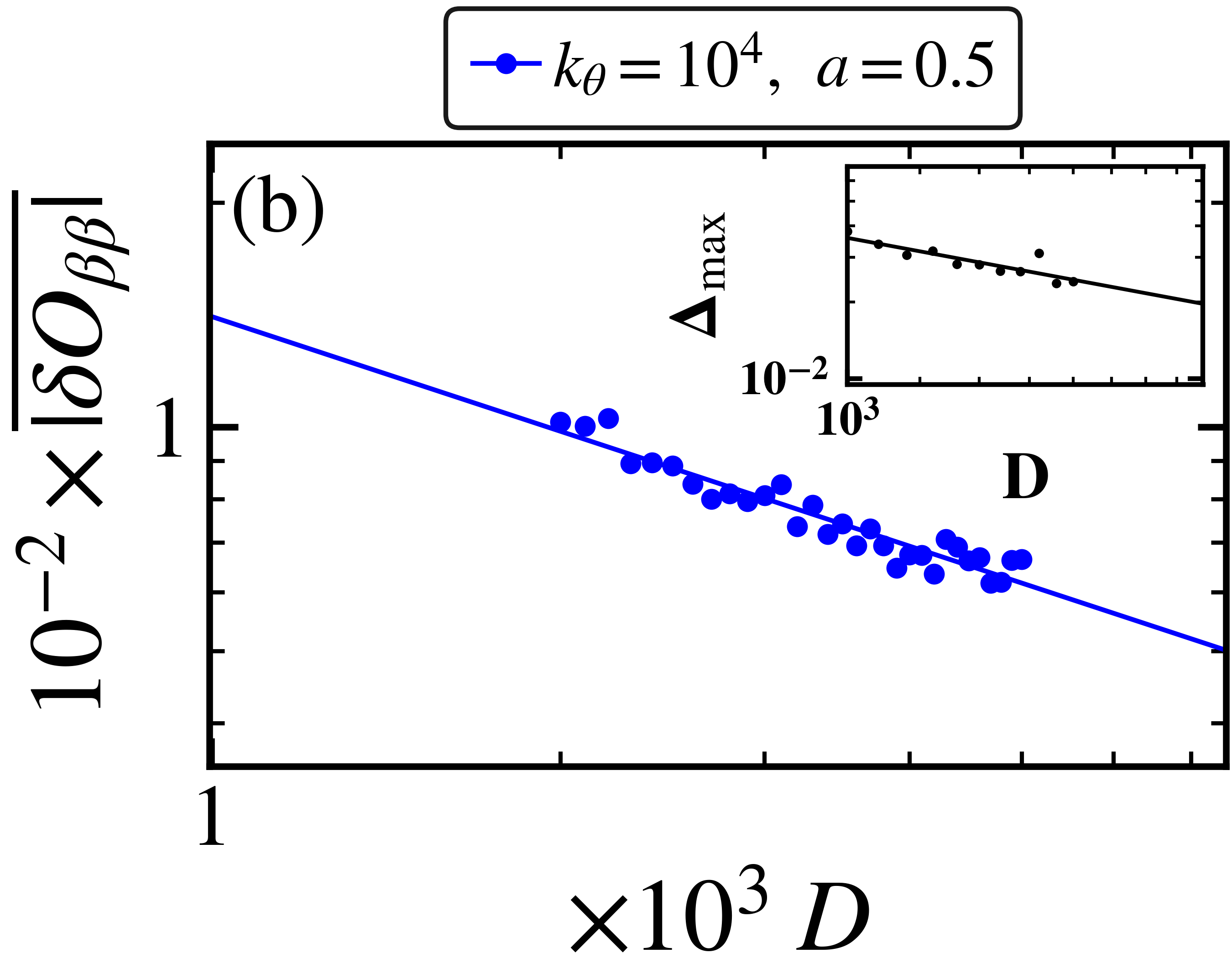}
    \caption{(a) Expectation value of normalized observable $J_z^2/j(j+1)$ in the eigenstate $|\phi_\alpha\rangle$ plotted versus $E_\alpha$ in DKT for $k_r = 1$ and $N = 2000$.~(b) Scaling of the average eigenstate-to-eigenstate fluctuations plotted against the parity-reduced Hilbert space dimensions $D$ for DKT at $k_r = 1$. For $k_\theta = 10^4$, fluctuations follow power-law $\overline{ | \delta O_{\beta\beta} | } \propto D^{-a}$. Inset: outlier deviations also follow $\Delta_{\max} \propto D^{-0.5}$.}\label{fig:eth-diag}
\end{figure}

\textit{Diagonal Matrix Elements --- }\label{sec:sub:diagonal}
To verify the ETH for diagonal elements, we plot the expectation value of the normalized observable $J_z^2/j(j+1)$ in the eigenstate $|\phi_\alpha\rangle$ versus the eigenenergy $E_\alpha$ for DKT at $k_r = 1$ (see Fig.~\ref{fig:eth-diag}~(a)). For small $k_\theta$, the expectation values show significant fluctuations around the microcanonical average (which is $1/3$ for $J_z^2/j(j+1)$). However, as $k_\theta$ increases, these fluctuations diminish, and the expectation values for all eigenstates converge towards the microcanonical average, showing \textit{strong} thermalization. 

To quantify this, we calculate the average eigenstate-to-eigenstate fluctuations as follows~\cite{leblond2019entanglement}: 
\begin{align}
    \widebar{\left|{\delta O}_{\alpha\alpha} \right|} = \widebar{\left|{O}_{\alpha, \alpha} - {O}_{\alpha+1,\alpha+1} \right|},
\end{align}
where the average is taken over all eigenstates. The scaling of $\widebar{\left|{(\delta J_z^2/j(j+1))}_{\alpha\alpha} \right|}$ versus parity-reduced Hilbert space dimensions $D$ at $k_r = 1$ is shown in Fig.~\ref{fig:eth-diag}~(b). For small (or close to zero) $k_\theta$, fluctuations are large and mostly remain independent of parity-reduced Hilbert space dimensions $D$. As we increase $k_\theta$, fluctuations start following power-law behavior $D^{-a}$ with $0 < a \leq 0.5$. For large values of $k_\theta$, these fluctuations reach $D^{-0.5}$, and the distribution approaches a Gaussian (see Fig.~\ref{supp:fig:eth-observable-distribution}~(a) in the supplementary material). As additional evidence, we compute an outlier deviation as follows~\cite{kim2014testing}:
\begin{align}
    \Delta_{\max} = \max_{\alpha} \big| O_{\alpha\alpha} - O_{\mathrm{mc}} \big|.
\end{align}
For the present analysis, we use $O_{\mathrm{mc}}=0.33$ (see Fig.~\ref{fig:eth-diag}~(a)). Our computations show that the outlier deviations also follow the power-law decay $\Delta_{\max} \sim D^{-0.5}$. These observations, together with fluctuations following a power-law $D^{-0.5}$, and a Gaussian distribution of diagonal matrix elements of an observable $\hat{O}$, demonstrate the \textit{strong} ETH~\cite{srednicki1994chaos,beugeling2014finite,d2016quantum,leblond2019entanglement}.

\begin{figure}
    \includegraphics[width=0.49\linewidth]{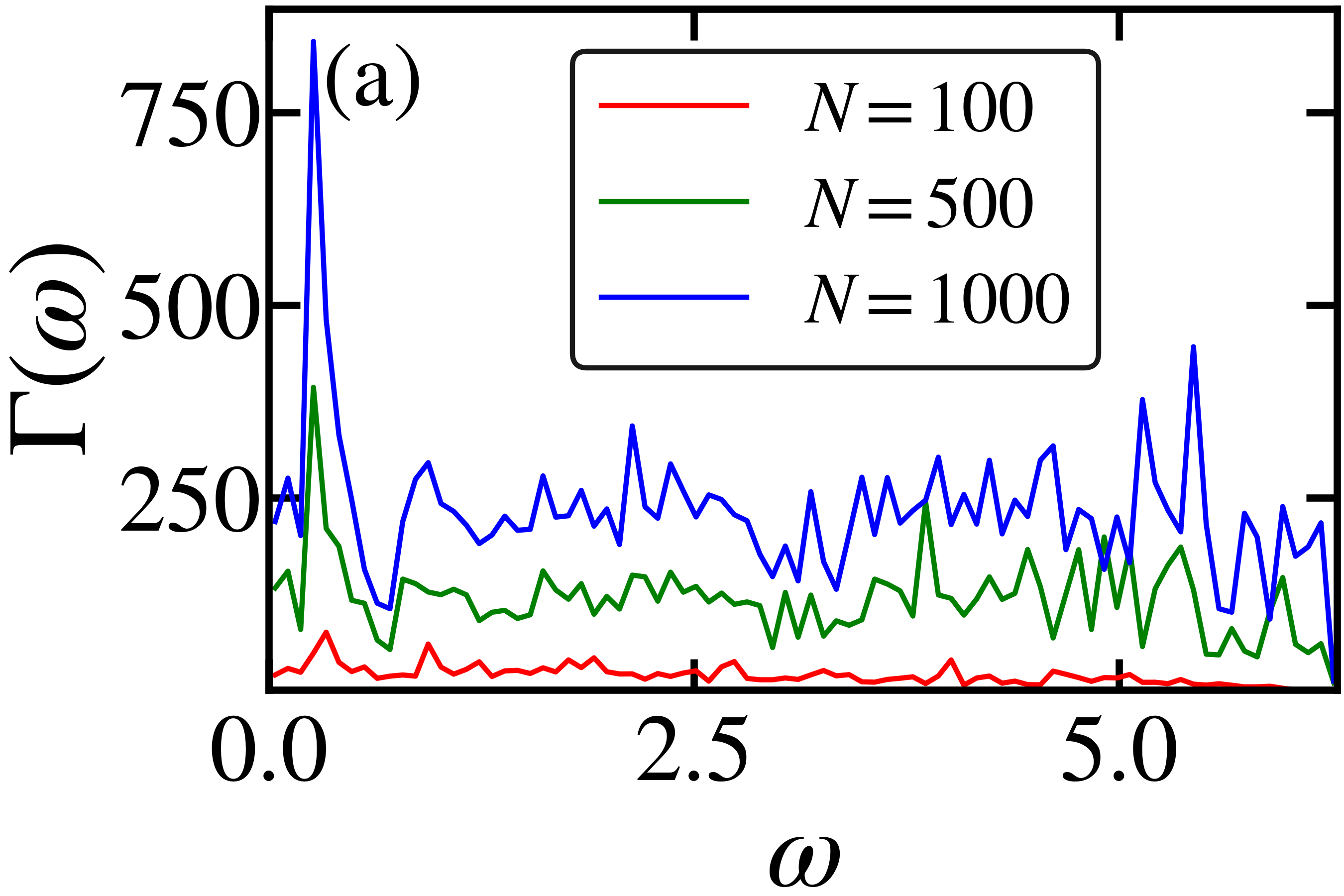}
    \includegraphics[width=0.49\linewidth]{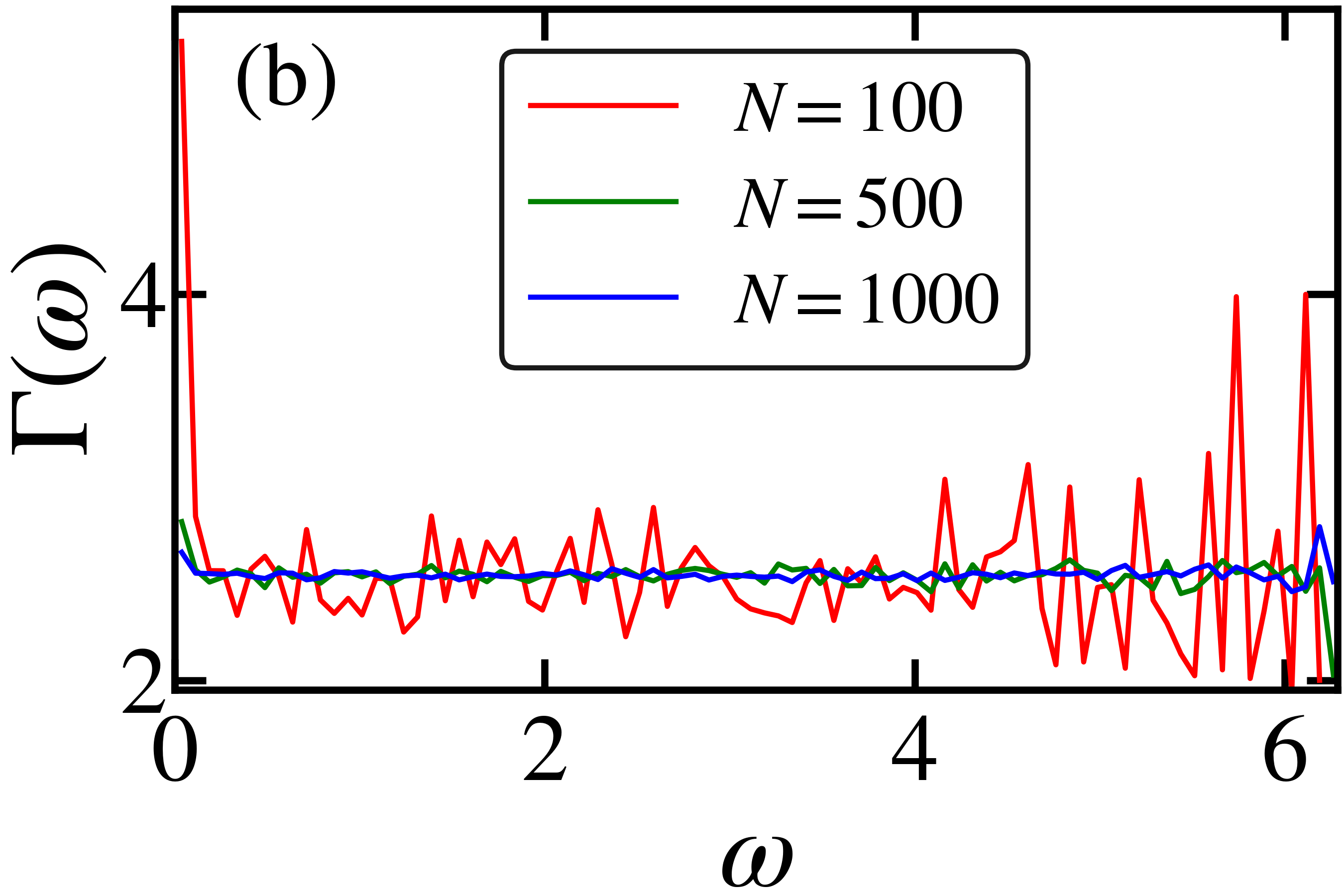}
    \caption{Gamma versus $\omega = | E_\alpha -E_\beta |$ for $k_r = 1$, (a) $k_\theta = 0$ and (b) $k_\theta = 10^4$. Growing fluctuations indicate a non-Gaussian nature for small $k_\theta$. Whereas, convergence to $\omega$-independent value at large $k_\theta$ signals Gaussian-like statistics.}\label{fig:eth-offdiag-gamma-vs-omega}
\end{figure}

\textit{Off-diagonal Matrix Elements --- }\label{sec:sub:off-diagonal}
For a consistency check, we analyze the off-diagonal elements of the same observable. Their non-Gaussian (or Gaussian) nature is given by~\cite{leblond2019entanglement,wang2024eigenstate}
\begin{align}
    \Gamma\!\left(\omega\right) = \frac{\widebar{|{O}_{\alpha\beta}|^2}}{{\widebar{|{O}_{\alpha\beta}|}}^2}.
\end{align}
For small $k_\theta$, the fluctuations $\Gamma \left(\omega\right)$ become more pronounced with increasing $N$, indicating highly non-Gaussian distribution (see Fig.~\ref{fig:eth-offdiag-gamma-vs-omega}~(a)). Whereas for large $k_\theta$, these fluctuations approach $\omega$-independent value, indicating a Gaussian-like distribution (see Fig.~\ref{fig:eth-offdiag-gamma-vs-omega}~(b)). Here, distribution is parameter-independent and $\sigma^2 \propto D^{-1}$, consistent with \textit{strong} ETH (see Fig.~\ref{supp:fig:eth-observable-distribution}~(b)).

\begin{figure}
    \includegraphics[width=0.495\linewidth]{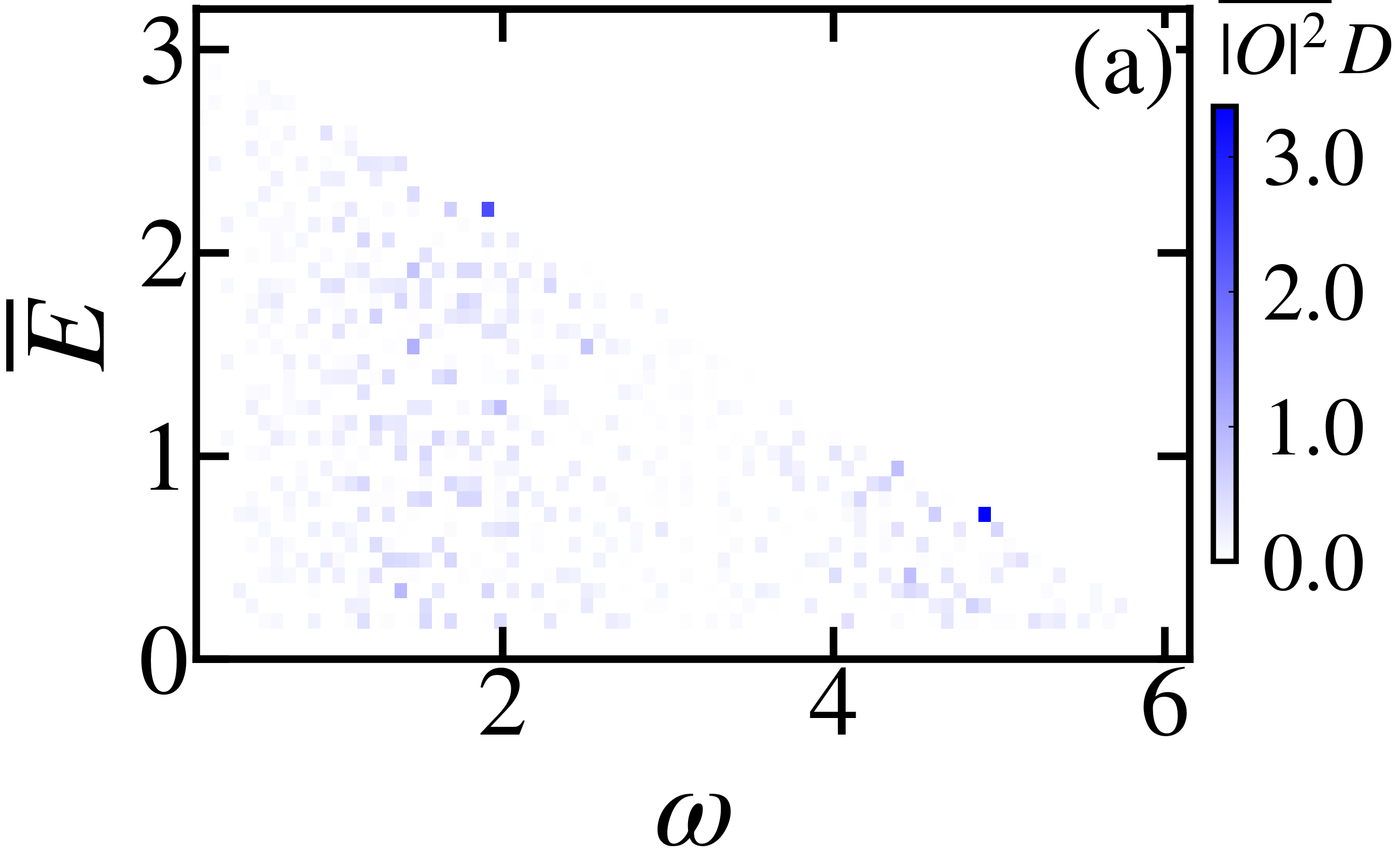}
    \includegraphics[width=0.495\linewidth]{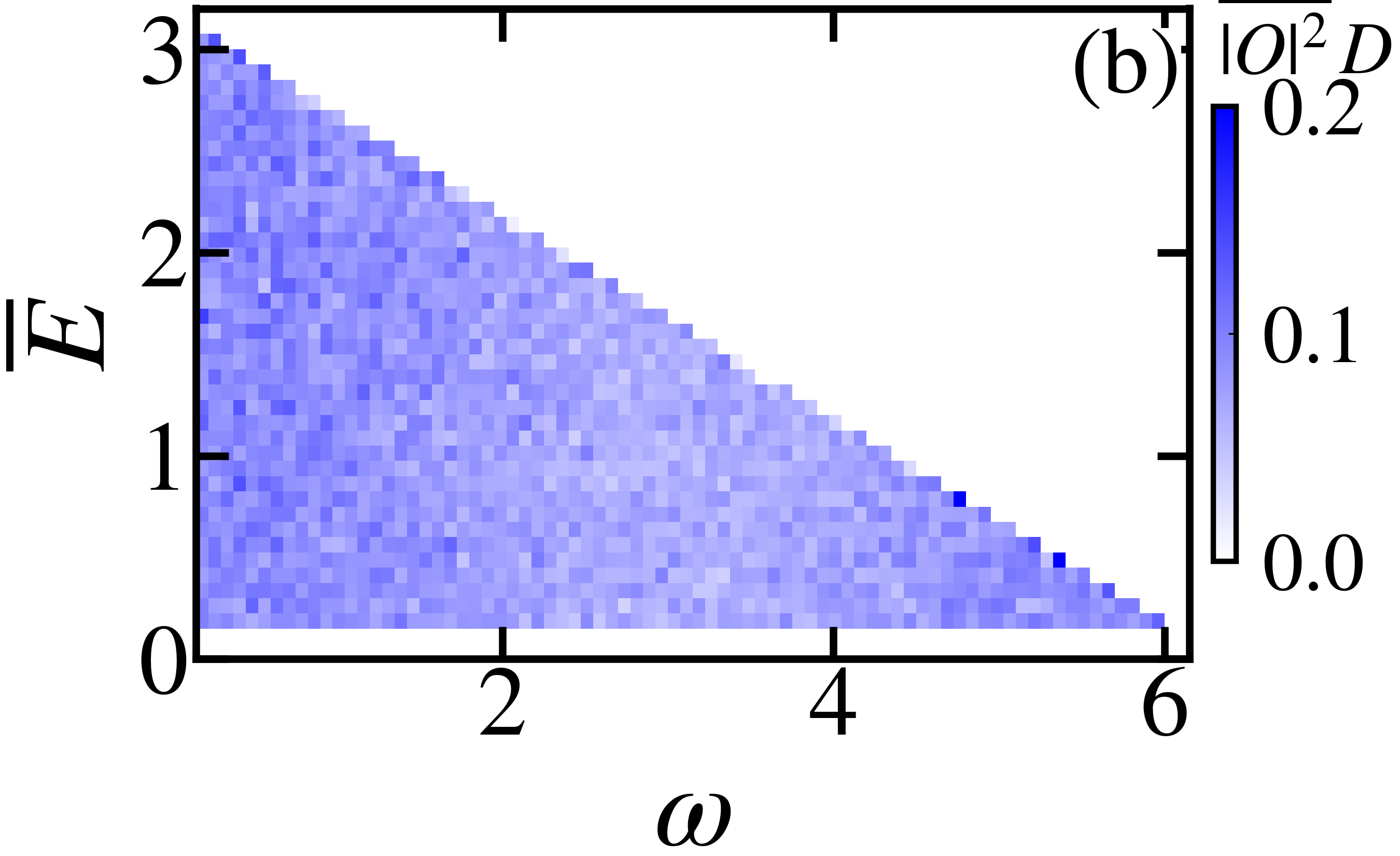}
    \caption{Distribution of smooth function ${\left|f_{\mathcal{O}}\left(\bar{E}, \omega\right)\right|}^2 = \overline{{|O|}^2} D$ plotted for off-diagonal elements of the DKT at $k_r = 1$ and $N=1000$ to observe the dependence of temporal fluctuations on the transformed kick strengths. Panel: (a) $k_\theta = 0$ and (b) $k_\theta = 10^4$.}\label{fig:smooth_function}
\end{figure}

Lastly, we analyze the smooth function ${\left|f_{\mathcal{O}}\left(\bar{E}, \omega\right)\right|}^2$ to know how the transition strength between a pair of eigenstates at specific energy varies with $k_\theta$ (see Fig.~\ref{fig:smooth_function}). For $k_\theta = 0$, the smooth function is highly localized, allowing only a few transitions (see Fig.~\ref{fig:smooth_function}~(a)). As $k_\theta$ is increased, the smooth function starts to spread across eigenstates and energies, allowing more transitions across the eigenstates and energies. For large values of $k_\theta$, the distribution becomes uniform over the entire region of validity (see Fig.~\ref{fig:smooth_function}~(b)). This uniform distribution indicates random-matrix-like behavior of $f_O(\bar{E},\omega)$ required for \textit{strong} ETH. As expected from the classical dynamics~\cite{berry1977level}, the corresponding \textit{strong} ETH domain, i.e., $(k_r \leq 1, k_\theta = 10^4)$, shows signatures of quantum integrability (see quantum-integrability Sec.~\ref{supp:sec:quantum-integrability} in the supplementary material).

\textit{Summary and Discussion --- }\label{sec:summary}
We have shown that \textit{strong} ETH emerges in a non-chaotic system with mean-ergodic dynamics. This is demonstrated using the DKT model, where the parameter $k_\theta$ induces mean-ergodic behavior without altering the degree of chaos. The resulting dynamics arise from the stretching and reorientation of phase-space structures around the trivial fixed points, leading to trajectories that spread over a large portion of the accessible phase-space. Consequently, observables converge to a common value over a longer duration, providing a classical basis for thermalization. In the quantum regime, the mechanism produces \textit{strong} ETH behavior. Thus, our analysis presents mean-ergodicity as a classical precursor to \textit{strong} ETH.

Since the QKT has been experimentally implemented using ${}^{133}\mathrm{Cs}$ atoms in the angular momentum $F=3$~\cite{chaudhary_quantum_signatures}, superconducting quantum processors~\cite{neill2016ergodic}, and quantum chaotic sensors~\cite{fiderer2018quantum}, these platforms can be naturally extended to realize the DKT. By probing the DKT in the regime $k_r \leq 1$ and large-$k_\theta$, one can experimentally access \textit{chaos-free thermalization}. It has been shown that quantum chaos leads to high entanglement growth~\cite{SilviaBridging2020,lakshminarayan2026quantum}, but it melts hardware and destroys the noninteracting qubit structure of the quantum computer~\cite{GeorgeotQuantumChaos2000,shepelyansky2001quantum,casati2003quantum,berke2022transmon}. Thus, in general, quantum chaos may not be suitable for quantum computational tasks. It is also known that entanglement is a useful resource for a quantum computer~\cite{HorodeckiQuantumEntanglement2009,mooney2021whole}. In this regard, our mechanism (\textit{chaos-free thermalization}) can be used to find potential candidates for an efficient quantum computer. We have shown that the mechanism is very well satisfied by DKT.

\textit{Acknowledgments --- }
The authors are grateful to the Department of Science and Technology (DST) for their generous financial support, making this research possible through sanctioned Project No. SR/FST/PSI/2017/5(C) to the Department of Physics of VNIT, Nagpur. The authors thank Dr.~Amit Anand and Dr.~Laura Foini for their valuable comments and suggestions on the manuscript.

    % f_{\mathrm{bad}} &= \frac{1}{D} \sum_{\alpha=1}^{D} \Theta\!\left( \left| O_{\alpha\alpha}-O_{\mathrm{mc}} \right| - \varepsilon \right) \;\, \text{and}\\

% -------------------------------
% BIBLIOGRAPHY (shared)
% -------------------------------
\bibliographystyle{apsrev4-2}
\bibliography{ref}

% -------------------------------
% SUPPLEMENT
% -------------------------------

% --- re-enable TOC ---
\makeatletter
\let\addcontentsline\addcontentslineOriginal
\makeatother

% --- Counter reset and custom formatting ---
\renewcommand{\thesection}{S\Roman{section}}
\renewcommand{\thesubsection}{\Alph{subsection}}
\renewcommand{\theequation}{S\arabic{equation}}
\renewcommand{\thefigure}{S\arabic{figure}}
\renewcommand{\thetable}{S\arabic{table}}

\setcounter{section}{0}
\setcounter{subsection}{0}
\setcounter{equation}{0}
\setcounter{figure}{0}
\setcounter{table}{0}

\renewcommand{\theHequation}{S\arabic{equation}}
\renewcommand{\theHfigure}{S\arabic{figure}}
\renewcommand{\theHtable}{S\arabic{table}}

% === 🧩 DEFINE CUSTOM SECTIONING COMMANDS ===

\newcommand{\suppsection}[2][]{%
  \refstepcounter{section}%
  \phantomsection%
  \section*{\texorpdfstring{S\Roman{section}\quad #2}{S\Roman{section} #2}}%
  \addcontentsline{toc}{section}{S\Roman{section}\quad #2}%
  \ifstrempty{#1}{}{%
    \label{#1}%
  }%
}

\newcommand{\suppsubsection}[2][]{%
  \refstepcounter{subsection}%
  \phantomsection%
  \subsection*{\texorpdfstring{\Alph{subsection}\quad #2}{\Alph{subsection} #2}}% <-- heading text without number
  \addcontentsline{toc}{subsection}{\thesubsection\quad #2}%
  \ifstrempty{#1}{}{%
    \label{#1}%
  }%
}

\clearpage
% \vspace*{1cm}
\onecolumngrid

\begin{center}
\textbf{\large Supplementary Material for\\ 
``\textit{Strong Eigenstate Thermalization from Mean-Ergodic Non-chaotic Dynamics}''}
\end{center}

\tableofcontents

\suppsection{Classical Dynamics}
In this section, we discuss some more phase-space trajectories, ergodic convergence using a function that does not depend on initial position, and the connection with the formal definition of mean-ergodicity, respectively. 

\suppsubsection{Mean-ergodic trajectories}\label{supp:sub:sec:trajectories}
To improve qualitative understanding of the mean-ergodic behavior, we discuss a couple of trajectories in a non-chaotic regime: $(\theta_0 = 2, \phi_0 = 2)$ and $(\theta_0 = \pi/2, \phi_0 = 0.1)$ (see Fig.~\ref{supp:fig:phase_space}). In the first case, blue period-8 orbits at $k_\theta = 0$ get stretched and reorient themselves with increasing $k_\theta$. In the extremely large $k_\theta$-limit, they occupy concentric annular shell-like regions. Similarly, the trajectory close to blue period-4 orbits occupies a nearby region but does not explore the outer region. These illustrations show that the trajectory explores only accessible phase space. Furthermore, it is not necessary for every trajectory to converge to the common value.
\begin{figure}[htbp]
    \includegraphics[width=0.48\linewidth]{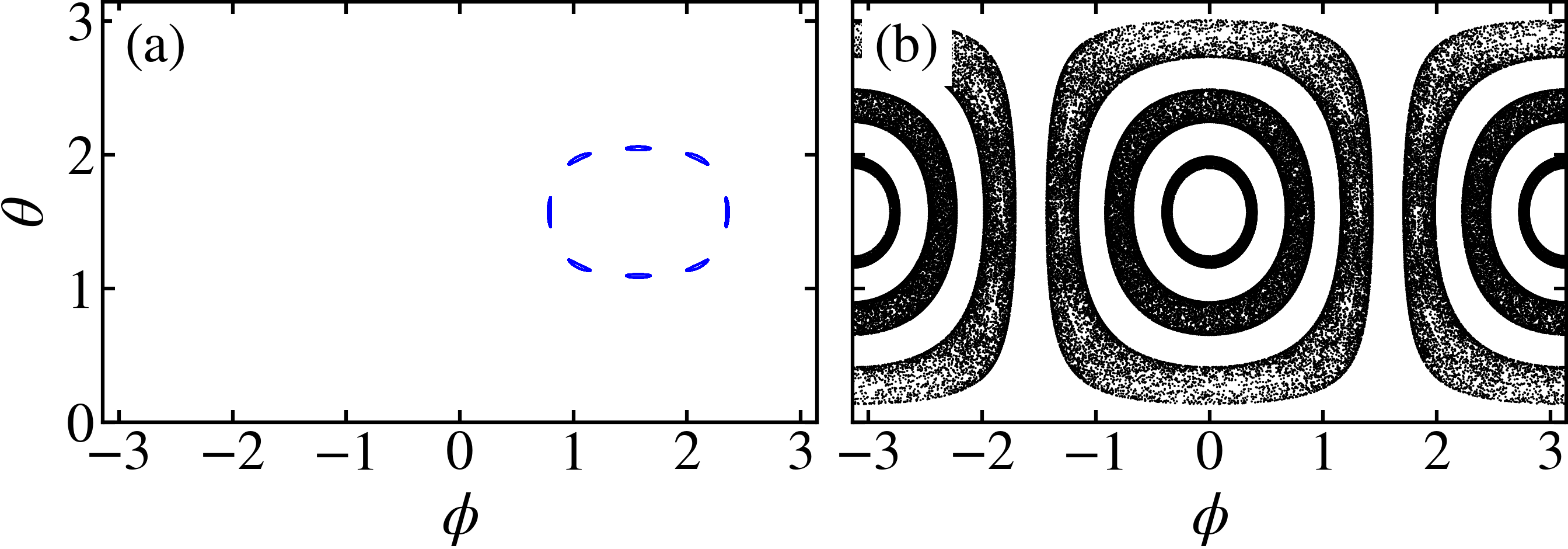}
    \includegraphics[width=0.48\linewidth]{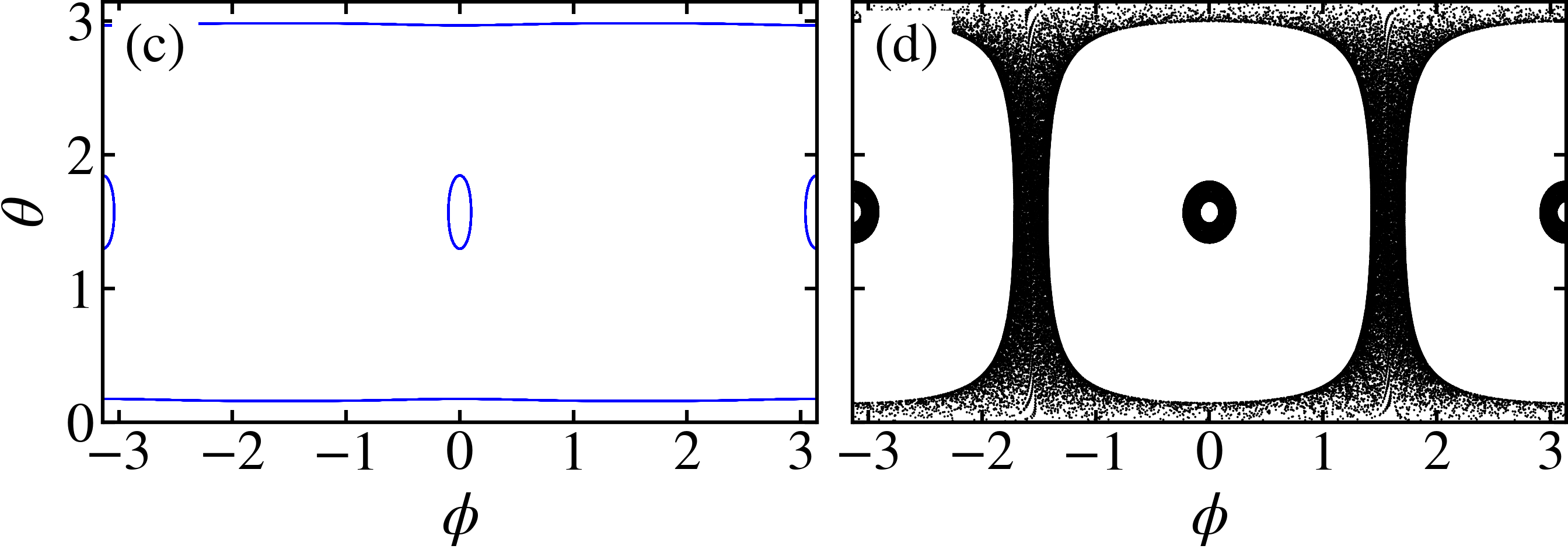}
    \caption{Phase-space evolution for two initial conditions evolved for $10^5$ kicks. Panels: (a) and (c) represent non-ergodic case $(k_r=1, k_\theta=0)$; (b) and (d) illustrate non-chaotic but mean-ergodic evolution for the case $(k_r=1, k_\theta=10^4)$.}\label{supp:fig:phase_space}
\end{figure}

\suppsubsection{Mean-ergodic convergence for a different observable}\label{supp:sub:sec:ergodicity}
To further test the robustness of mean-ergodic convergence discussed in the main text, we consider an additional observable $ f(\theta, \phi) = \sin(\theta)\sin(\phi) $. Unlike the trajectory-length observable $l_\tau(\mathbf{X}_\tau,\mathbf{X}_0)$ used in the main text, this function does not explicitly depend on the initial condition, allowing us to test ergodic convergence for a purely phase-space observable. We compute the ensemble fluctuations $\operatorname{Var}_\tau(f)$ for different dynamical regimes (see Fig.~\ref{supp:fig:ergodicity}). For the non-ergodic case $(k_r = 1, k_\theta = 0)$, the fluctuations saturate to a non-zero value at long times, indicating persistent dependence on initial conditions. In contrast, all other cases $(k_r = 1, k_\theta = 10^4)$, $(k_r = 3, k_\theta = 0)$, and $(k_r = 5, k_\theta = 0)$ show decay of fluctuations consistent with a power law $\operatorname{Var}_\tau(f) \propto \tau^{-\alpha}$, with $\alpha \approx 1$ for the strongly chaotic regime and slower decay in the weakly chaotic and non-chaotic cases. Notably, the case $(k_r = 1, k_\theta = 10^4)$ is non-chaotic yet still exhibits decay of fluctuations toward zero, demonstrating mean-ergodic convergence despite the absence of chaos. This contrasts with $(k_r = 3, k_\theta = 0)$ and $(k_r = 5, k_\theta = 0)$, where convergence is driven by chaotic dynamics.
\begin{figure}
    \includegraphics[width=0.3\linewidth]{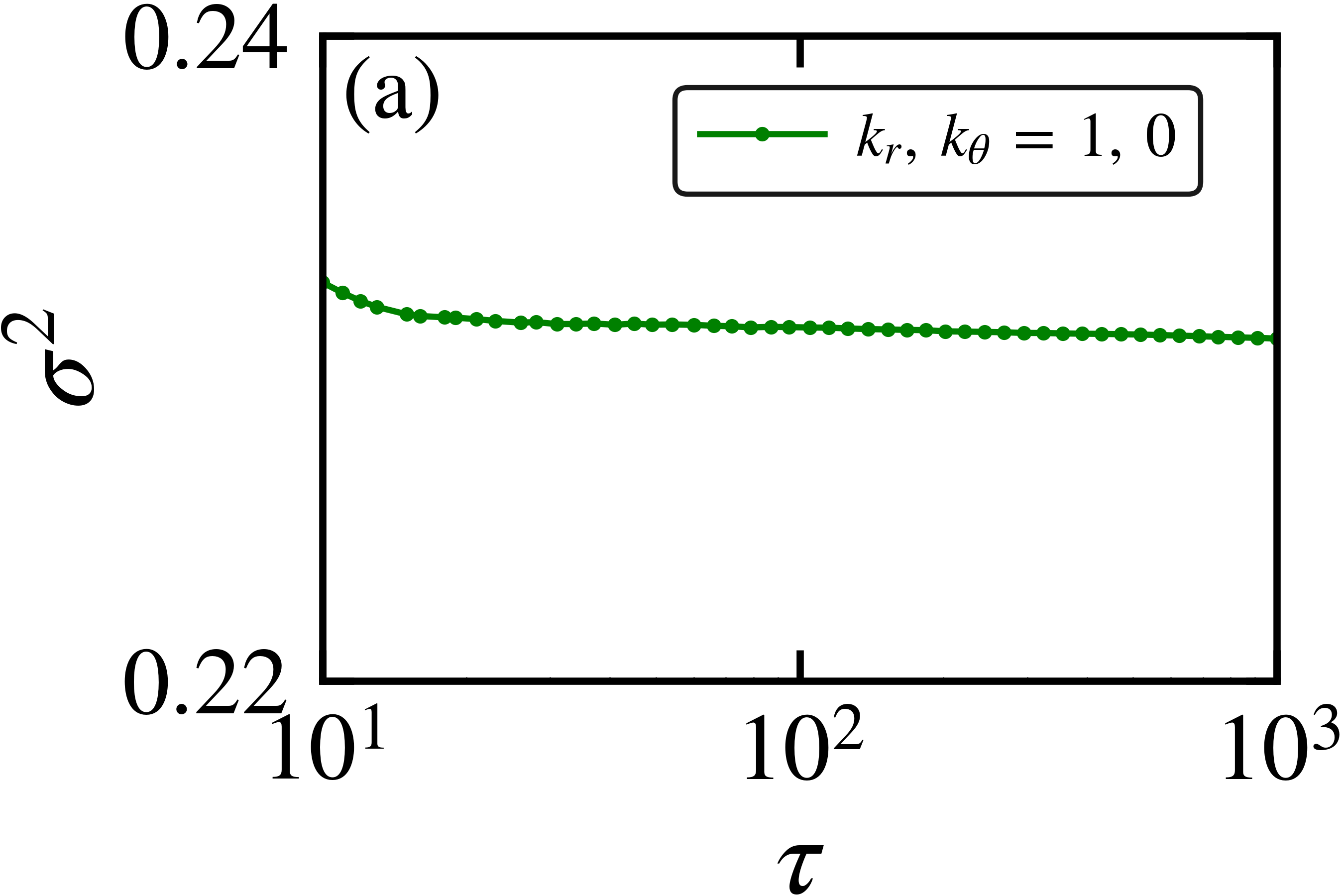}
    \includegraphics[width=0.3\linewidth]{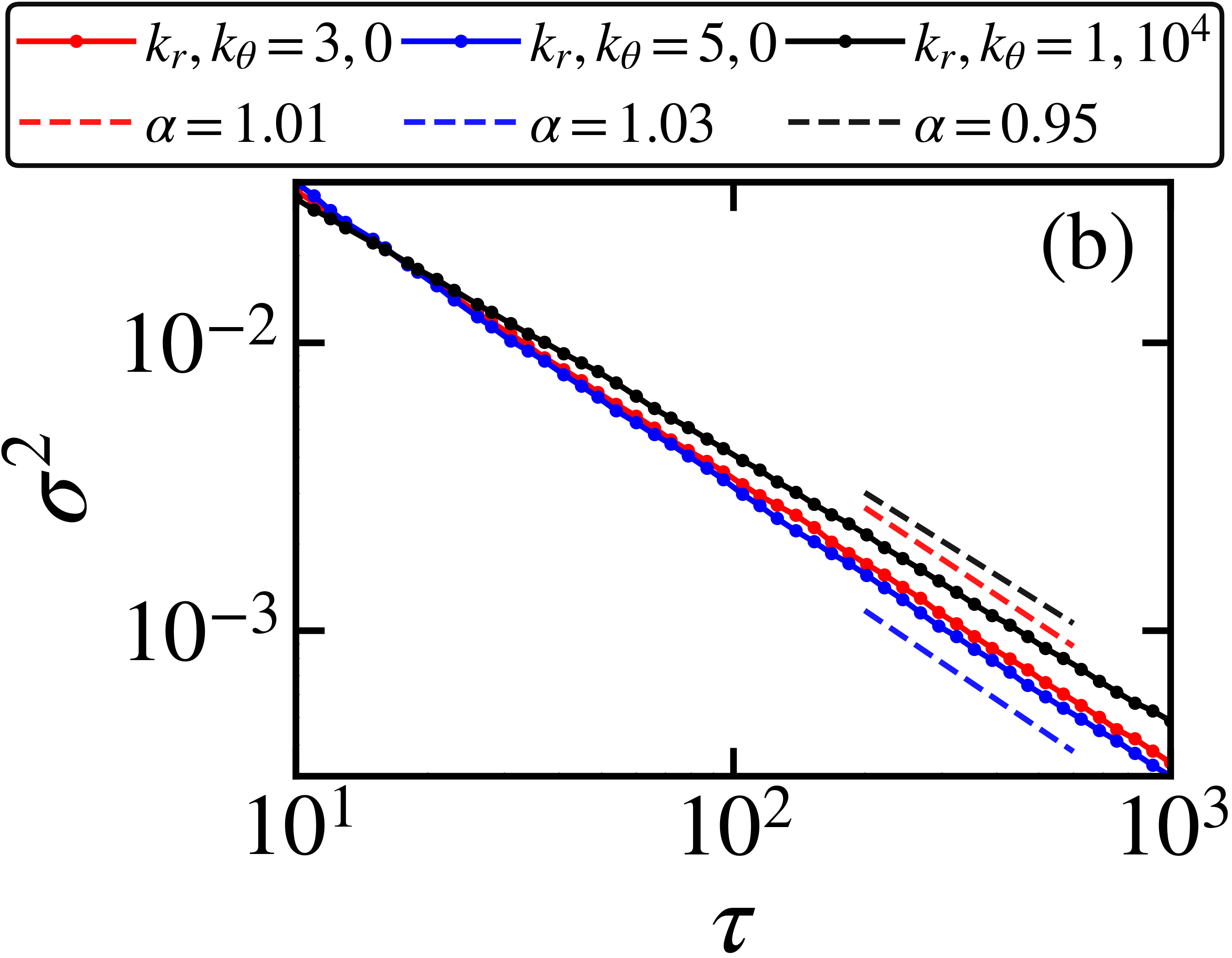}
    \caption{Fluctuations $\operatorname{Var}_\tau(f)$ for $f(\theta, \phi) = \sin(\theta)\sin(\phi)$ fitted with the scaling $\operatorname{Var}_\tau(f) \propto \tau^{-\alpha}$. (a) Non-ergodic regime: $k_r=1$, $k_\theta=0$. (b) Mean-ergodic regimes: chaotic cases $(k_r = 3, k_\theta = 0)$ and $(k_r = 5, k_\theta = 0)$, and non-chaotic case $(k_r = 1, k_\theta = 10^4)$.}\label{supp:fig:ergodicity}
\end{figure}

\suppsubsection{Connection with the formal definition of mean-ergodicity}\label{supp:sub:sec:analogy}
To connect Eq.~\eqref{eq:variance} used in the main text with the formal definition of mean-ergodicity, we recall the von Neumann's mean-ergodic theorem~\cite{neumann1932proof,cornfeld2012ergodic}. For a measure-preserving dynamical system, the evolution of observables is described by the Koopman operator $U$, acting as $(U^n A)(\mathbf{X}_0) = A(T^n \mathbf{X}_0)$. The finite-time ergodic average of an observable can, therefore, be written as follows:
\begin{align}
    \bar{A}_\tau(\mathbf{X}_0) = \frac{1}{\tau}\sum_{n=0}^{\tau-1} U^n A(\mathbf{X}_0).
\end{align}
This operation is defined and discussed in the von Neumann's mean-ergodic theorem~\cite{neumann1932proof,cornfeld2012ergodic}. The theorem states that this time-averaged observable converges in the mean-square sense as,
\begin{align}
    \left\| \bar{A}_\tau - \mathcal{P}A \right\|_2 \to 0 \quad \text{as } \tau \to \infty,
\end{align}
where $\mathcal{P}$ projects onto invariant observables, reducing to the phase-space average $\langle A \rangle$ in the ergodic case. To make the connection with numerics explicit, we note that our fluctuation measure is the ensemble variance of these finite-time Koopman averages:
\begin{align}
    \mathrm{Var}_{\mathrm{ens}}(\bar{A}_\tau) = \left\langle \bar{A}_\tau^2 \right\rangle_{\mathrm{ens}} - \left\langle \bar{A}_\tau \right\rangle_{\mathrm{ens}}^2.
\end{align}
Expanding $\bar{A}_\tau^2$ shows that this quantity corresponds to correlations of Koopman iterates,
\begin{align}
    \bar{A}_\tau^2 = \frac{1}{\tau^2}\sum_{n,m=0}^{\tau-1} U^n A \cdot U^m A,
\end{align}
so that the variance probes the decay of temporal correlations encoded in the Koopman evolution. In the limit of large ensemble size, this variance approximates the $L^2$ norm with respect to the invariant measure appearing in the von Neumann theorem~\cite{petersen1995ergodic}:
\begin{align}
    \mathrm{Var}_{\mathrm{ens}}(\bar{A}_\tau) \;\longrightarrow\; \left\| \frac{1}{\tau}\sum_{n=0}^{\tau-1} U^n A - \langle A \rangle \right\|_{2}.
\end{align}
Therefore, the decay of $\operatorname{Var}_\tau(A)$ to zero provides a direct numerical signature of mean-ergodic convergence in the sense of von Neumann.

\suppsection{Quantum Dynamics}\label{supp:sec:quantum_dynamics}
In this section, we discuss von Neumann entropy results, distribution of diagonal and off-diagonal matrix elements, and signatures of quantum-integrability in the domain of interest.

\suppsubsection{Dynamical entropy}\label{supp:sub:sec:vne}
We consider $2j$ spin-half particles with all-to-all interactions and total spin $j$. The Floquet operator can be rewritten in terms of these $N=2j$ number of spin-half particles using total angular momentum $\sum_{l=1}^{2j} \sigma_l^{x,y,z}/2$~\cite{milburn1999simulating,wang2004entanglement,sharma2024exactly,purohit2025double} as follows:  
\begin{align}\label{Floquet}
    \hat{\mathcal{U}} &= \exp\!\left(-i \frac{k'}{4j} \sum_{l' < l = 1}^{2j} \sigma_{l'}^x \sigma_l^x \right) \exp\!\left(-i \frac{k}{4j} \sum_{l' < l = 1}^{2j} \sigma_{l'}^z \sigma_l^z \right) \exp\!\left(-i \frac{\pi}{4} \sum_{l=1}^{2j} \sigma_l^y\right).
\end{align}
The standard $SU(2)$ coherent states~\cite{glauber1976superradiant, puri2001mathematical} (in qubit basis) are considered as initial states:  
\begin{align}\label{Eq:generalstate}
    |\theta_0, \phi_0\rangle = \otimes^{2j}\!\left[\cos\left(\frac{\theta_0}{2}\right) | 0\rangle + e^{-i\phi_0}\sin\left(\frac{\theta_0}{2}\right)| 1 \rangle\right].
\end{align}
To improve computational efficiency, we expressed them in the $\lbrace |j,m\rangle \rbrace$ basis using Eq.~(12) of~\cite{Bandyopadhyay2004}.

To visualize thermalization, we consider the long-time-averaged von Neumann entropy $S$ corresponding to the single-particle reduced density matrix $\rho_1(n)$. It is defined as follows:
\begin{align}
    \langle S(k_r, k_\theta) \rangle = \lim_{t\to\infty} \frac{1}{t} \sum_{n=0}^{t-1} S(n, k_r, k_\theta) \; \text{ where }\; S(n, k_r, k_\theta) = - \text{tr}\left( \rho_1 \log_2 \rho_1 \right).
\end{align}
The long-time-averaged von Neumann entropy $\langle S \rangle$ landscape for $k_r = 1$ and $k_\theta = 0$ reveals fine-grained phase-space structures (see Fig.~\ref{supp:fig:vn_entropy}(a)). The blue low entanglement regions indicate trivial fixed points with red chaotic borders corresponding to the period-4 cycles. The entanglement increases with $k_\theta$ (also see Fig.~(25),~\cite{purohit2025double}). For large values of $k_\theta$, the von Neumann entropy for almost the entire region gets saturated to its maximum value (see Fig.~\ref{supp:fig:vn_entropy}(b)), providing evidence consistent with ETH-like thermalization.
\begin{figure}
    \centering
    \includegraphics[width=0.5\linewidth]{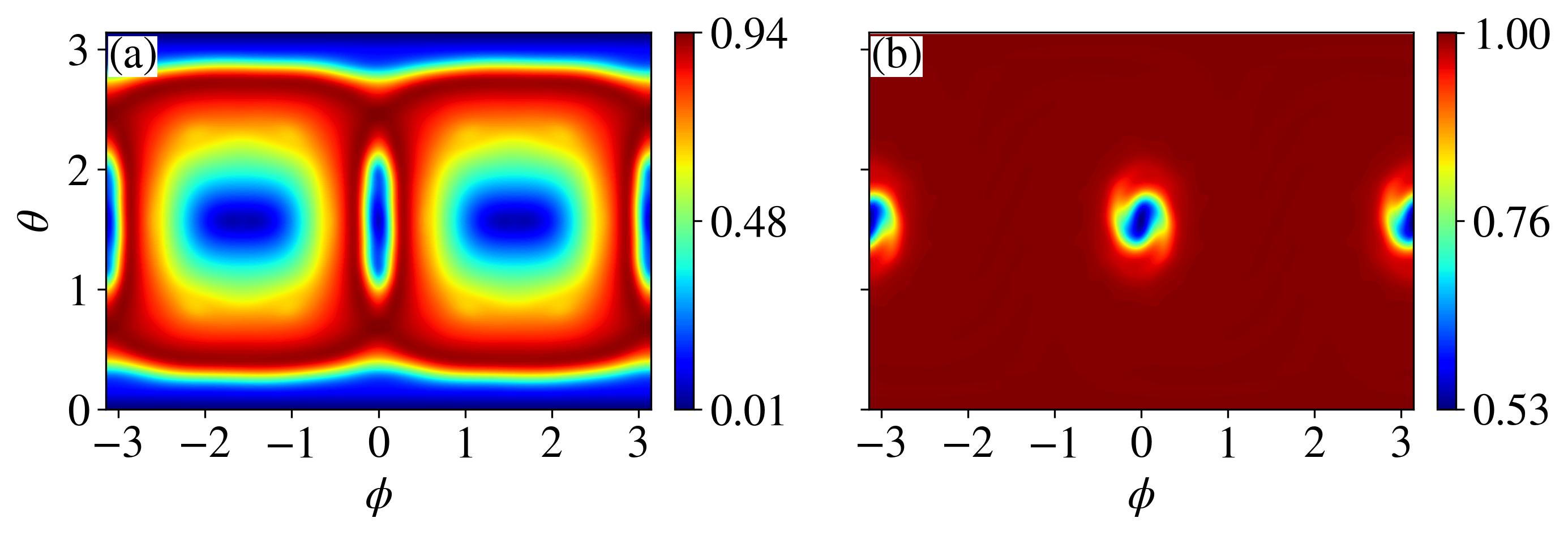}
    \caption{The long-time-averaged von Neumann entropy for the state represented by the single-particle reduced density matrix $\rho_1(n)$ having a grid of $200 \times 200$ initial conditions evolved for 1000 kicks. Here, kick strength $k_r=1$, $j=75.5$, (a) $k_\theta = 0$ and (b) $k_\theta = 10^4$.}\label{supp:fig:vn_entropy}
\end{figure}

\suppsubsection{Observable distribution}\label{supp:sub:sec:observable-distribution}
In this section, we analyze the distribution of diagonal and off-diagonal matrix elements of the observable $O = J_z^2 / j(j+1)$ for different kicking strengths (see Fig.~\ref{supp:fig:eth-observable-distribution}). The distribution of diagonal matrix elements evolves from non-Gaussian to Gaussian-like as $k_\theta$ increases. This indicates suppression of large eigenstate-to-eigenstate fluctuations. This behavior is consistent with the emergence of strong ETH in the large-$k_\theta$ regime. In contrast, the distribution of off-diagonal matrix elements shows no qualitative distinction between non-ergodic, mean-ergodic, and chaotic regimes. This insensitivity suggests that off-diagonal ETH indicators are considerably less sensitive to the dynamical phase than diagonal ETH indicators associated with strong ETH. One of the possible reason have been investigated by Ref.~\cite{huang2025trade}. 
\begin{figure}
    \centering
    \includegraphics[width=0.35\linewidth]{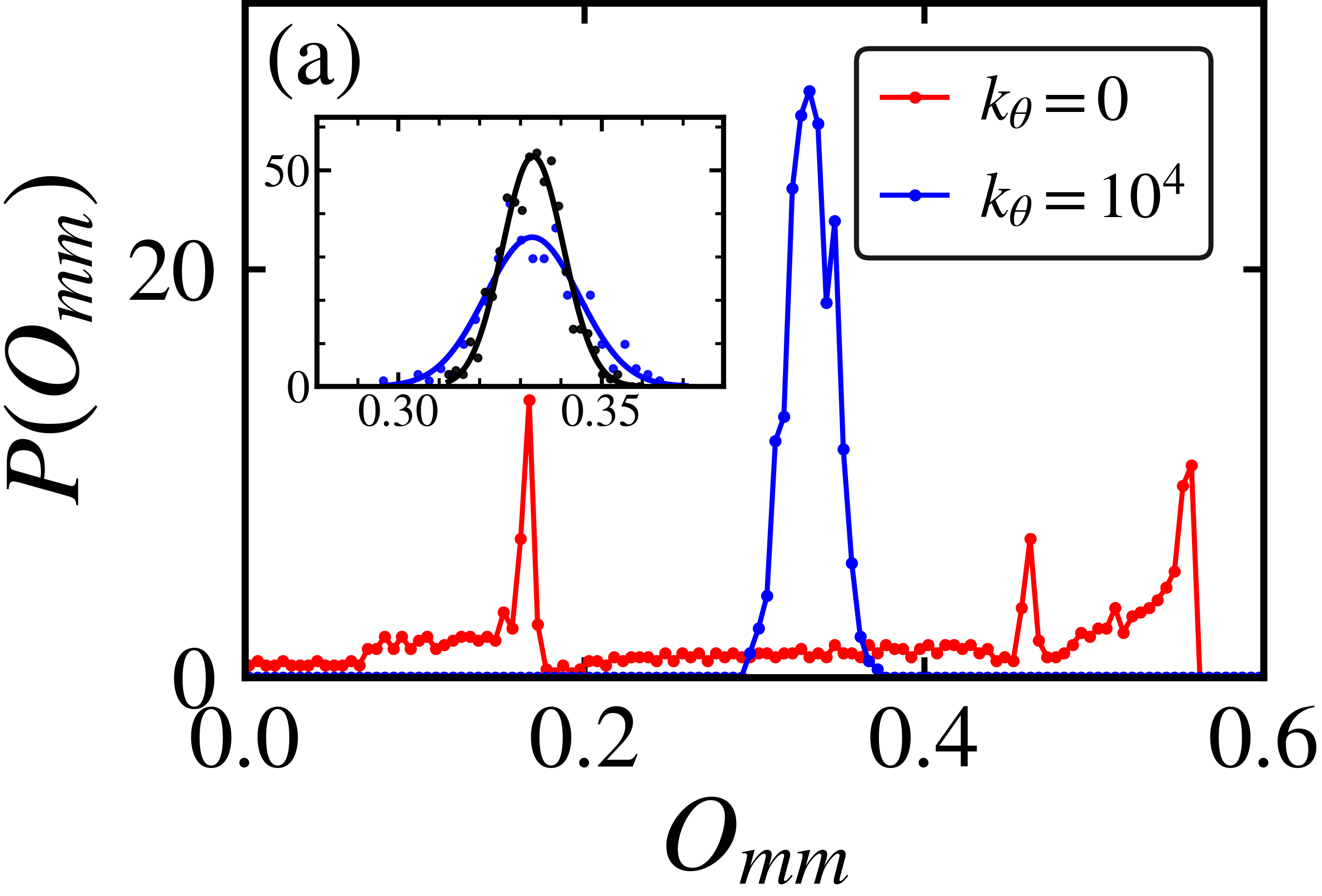}
    \includegraphics[width=0.35\linewidth]{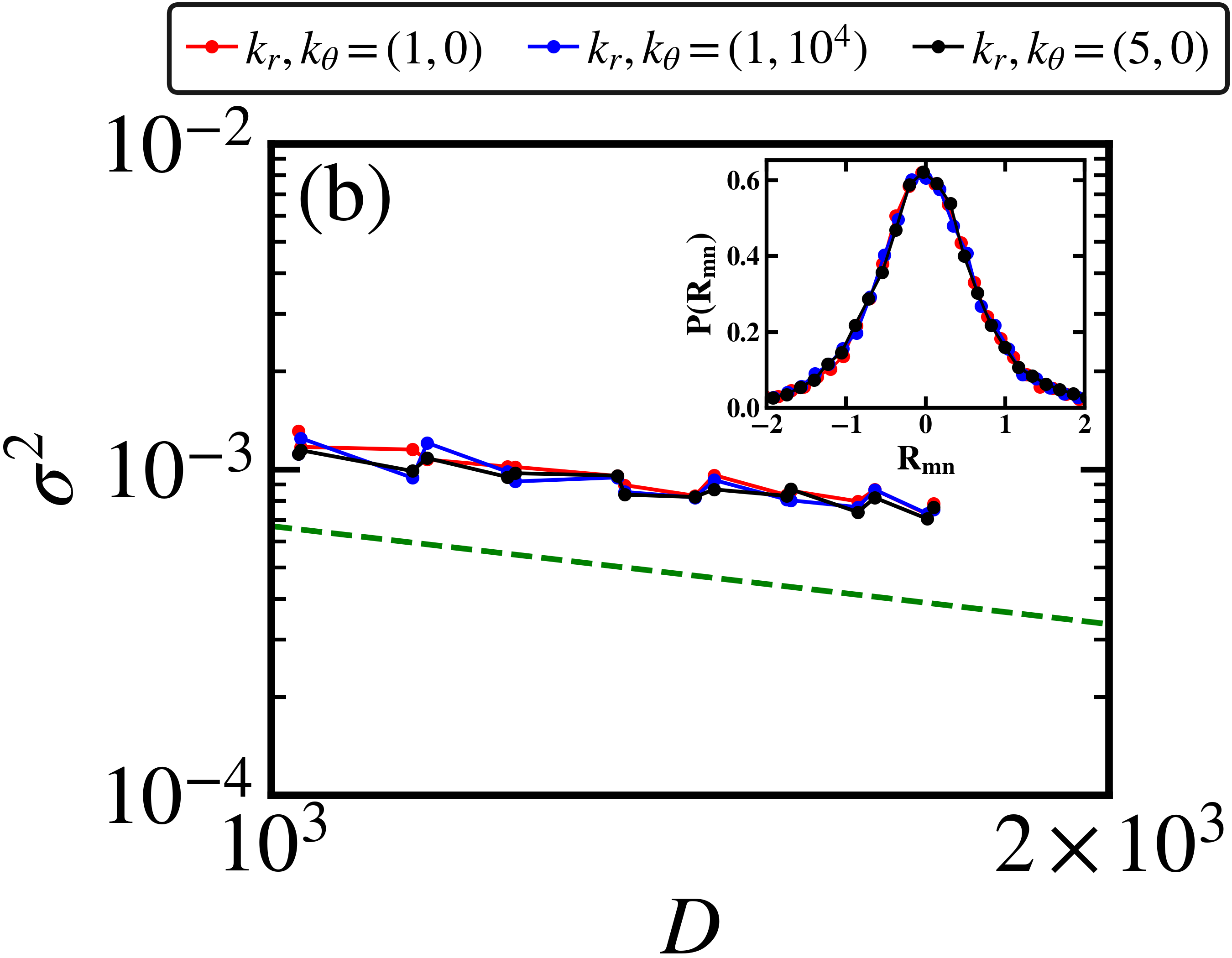}
    \caption{(a) Histogram for diagonal elements of $J_z^2/j(j+1)$ plotted for $j=1000$, $k_r = 1$. The case $k_\theta = 0$ represented by red and $k_\theta = 10^4$ is illustrated in blue. Inset: Gaussian distribution approaches Dirac delta distribution with increase in system size from blue, $N=1500$ to black, $N=3500$.~(b) Scaling of the off-diagonal variance $\sigma^2$ with the parity-resolved Hilbert-space dimension $D$ for system sizes $N=2000$--$3500$. The variance is computed from off-diagonal matrix elements within a fixed quasienergy window and parity sector. For the non-chaotic cases $(k_r,k_\theta)=(1,0)$ and $(1,10^4)$, and the chaotic case $(5,0)$, the variance shows the ETH scaling $\sigma^2 \propto D^{-1}$ (green dashed line). Inset: Histogram of the normalized off-diagonal matrix elements $R_{mn}=O_{mn}/\sigma$ after parity resolution.}\label{supp:fig:eth-observable-distribution}
\end{figure}
%  (also see Fig.~8 of Ref.~\cite{wang2024eigenstate})
% leptokurtic distribution
% (b) Histogram for off-diagonal elements $R_{mn} = O_{mn}/\sigma$ plotted by resolving parity for $j=2000$. The variance $\sigma^2$ is computed over off-diagonal elements within a fixed quasienergy window and symmetry sector. We consider two non-chaotic cases: $(k_r = 1, k_\theta = 0)$, $(k_r = 1, k_\theta = 10^4)$ and one chaotic case $(k_r = 5, k_\theta = 0)$. Inset: The variance $\sigma^2 \propto D^{-1}$ for all three cases.

\begin{figure}
    \centering
    \includegraphics[width=\linewidth]{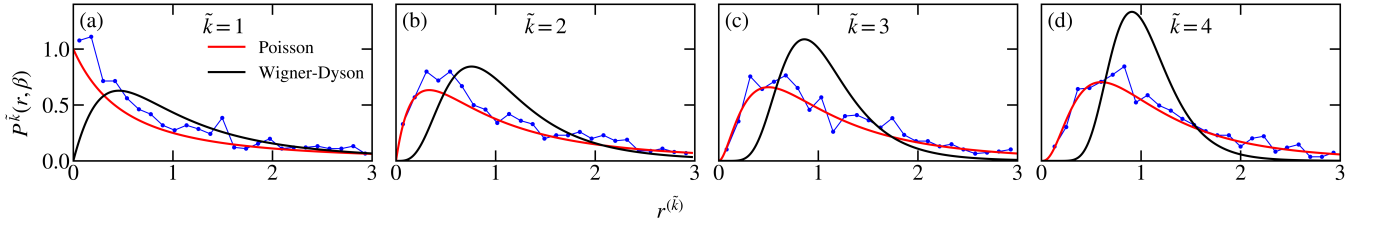}
    \caption{The probability distribution \(P^{\tilde{k}}(r, \beta)\) versus \(r^{(\tilde{k})}\) plotted for the DKT with \(p = \pi/2\),  \(k_r = 1 \), \( k_\theta = 10^4 \), and \(j = 2000.5\). Panels (a)-(d) correspond to order of $\tilde{k}$ from 1 to 4 respectively.}\label{supp:fig:level-spacing}
\end{figure}

\suppsubsection{Quantum-integrability for large \(k_\theta\)}\label{supp:sec:quantum-integrability}
The quantum integrability of the DKT in the domain $k_r\leq 1$ and small $k_\theta$ is already shown in Ref.~\cite{purohit2025double}. To support quantum integrability for a large $k_\theta$ domain (keeping $k_r \leq 1$), we compute the higher-order spacing ratio statistics. Here, the $\tilde{k}^{\text{th}}$ order non-overlapping spacing ratio is given by~\cite{bhosale2018scaling,harshini2020symmetry,bhosale2028higher}
\begin{align}
    r^{(\tilde{k})} = \frac{E_{i+2\tilde{k}} - E_{i+\tilde{k}}}{E_{i+\tilde{k}} - E_i},
\end{align}
where $\lbrace E_i\rbrace$ represent quasi-energies of the Floquet operator $\mathcal{U}$. Since the nearest-neighbor statistics may sometimes give misleading results in the presence of some unresolved symmetries in the system. The higher-order spacing ratio statistics (HOSRS) are, thus, useful in determing true (integrable or chaotic) nature in most of the cases~\cite{tkocz2012tensor,bhosale2021superposition,giraud2022probing}. For this reason, we employ HOSRS to avoid such possible misleading results. The $\tilde{k}^{\text{th}}$ order spacing ratio distribution for uncorrelated energy levels is given as follows~\cite{harshini2020symmetry}:
\begin{align}
    P_P^{(\tilde{k})}(r) &= \frac{\Gamma(2\tilde{k})\, r^{\tilde{k}-1}}{\Gamma(\tilde{k})^2 (1+r)^{2\tilde{k}}}.
\end{align}
The results are shown in Fig.~\ref{supp:fig:level-spacing}. They agree with the Poisson distribution supporting quantum integrability for the said parameters.

We have also studied long-range correlations using the spectral form factor (SFF) to further support the quantum integrability~\cite{brezin1997spectral,berry1985semiclassical,Haakebook,dong2025measuring}. It is defined as follows:
\begin{eqnarray}
    K(t,N)&=&\big \langle |\mbox{Tr}[\mathcal{\hat{U}}(t)]|^2 \big\rangle=\Big \langle \sum_{a,b}\exp \left(i \left(E_a-E_b\right)t\right) \Big\rangle.
\end{eqnarray}
The analytical expression of SFF for Poisson distribution is given as follows:
\begin{equation}\label{eq:sff-int}
    K^P(t)=N+\frac{2}{{(\mu t)}^2}-\frac{{(1+i \mu t)}^{1-N}+{(1-i \mu t)}^{1-N}}{{(\mu t)}^2}.
\end{equation}
\begin{figure}
    \includegraphics[width=0.7\linewidth]{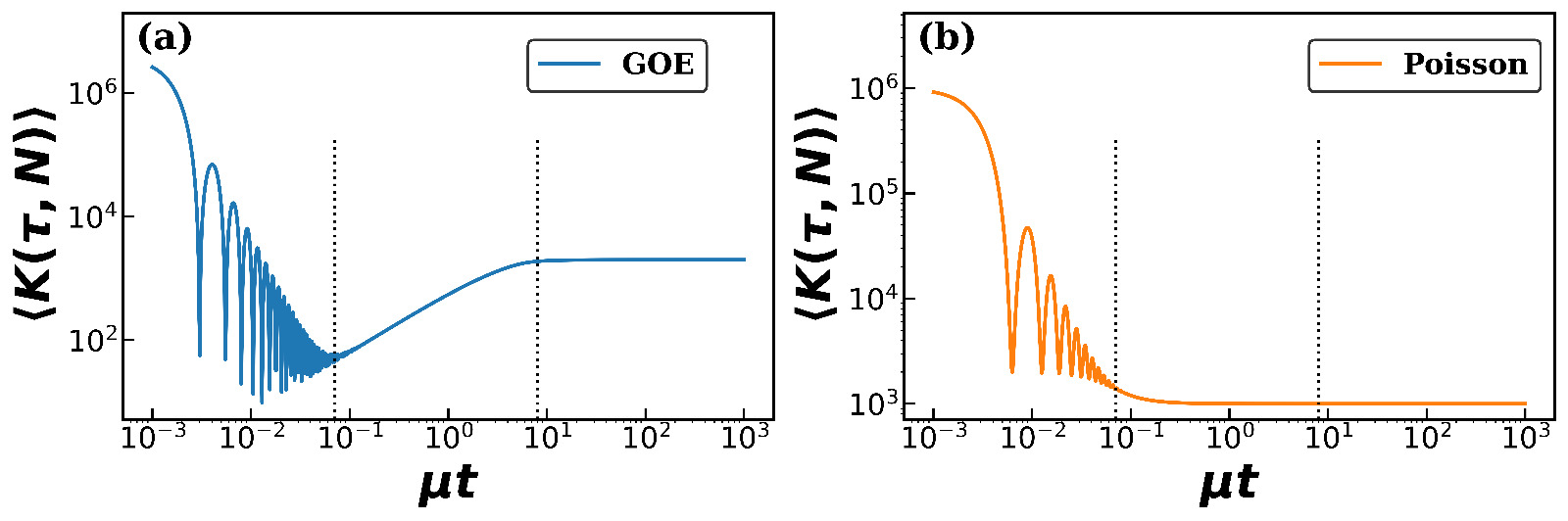}
    \caption{The SFF corresponding to $N = 1000$ is shown in Panel: (a) GOE from Eq.~\eqref{eq:sff-goe} (a) Poisson from Eq.~\eqref{eq:sff-int}.}\label{supp:fig:sff-ideal}
\end{figure}
\begin{figure}
    \includegraphics[width=0.35\linewidth]{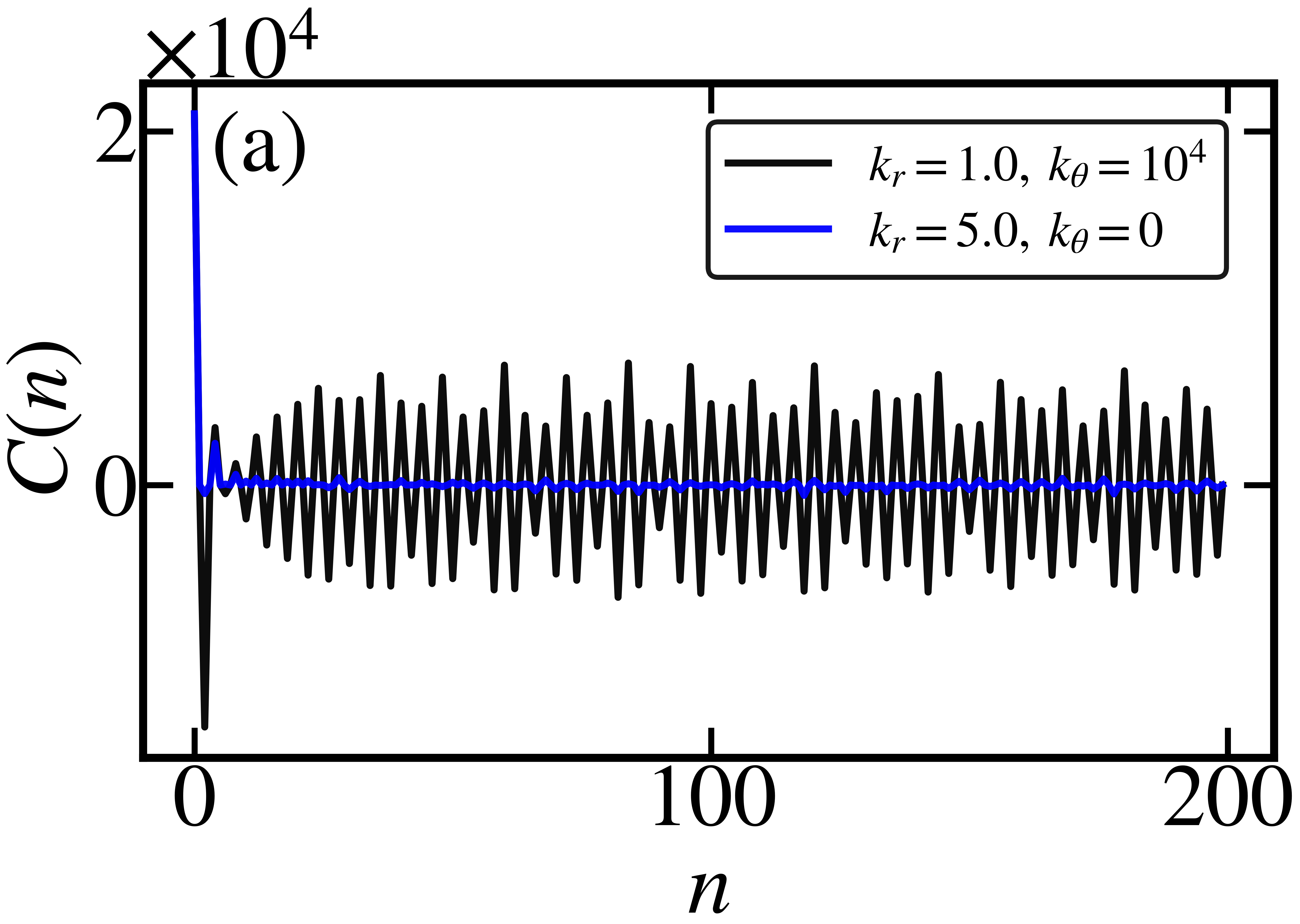}
    \includegraphics[width=0.35\linewidth]{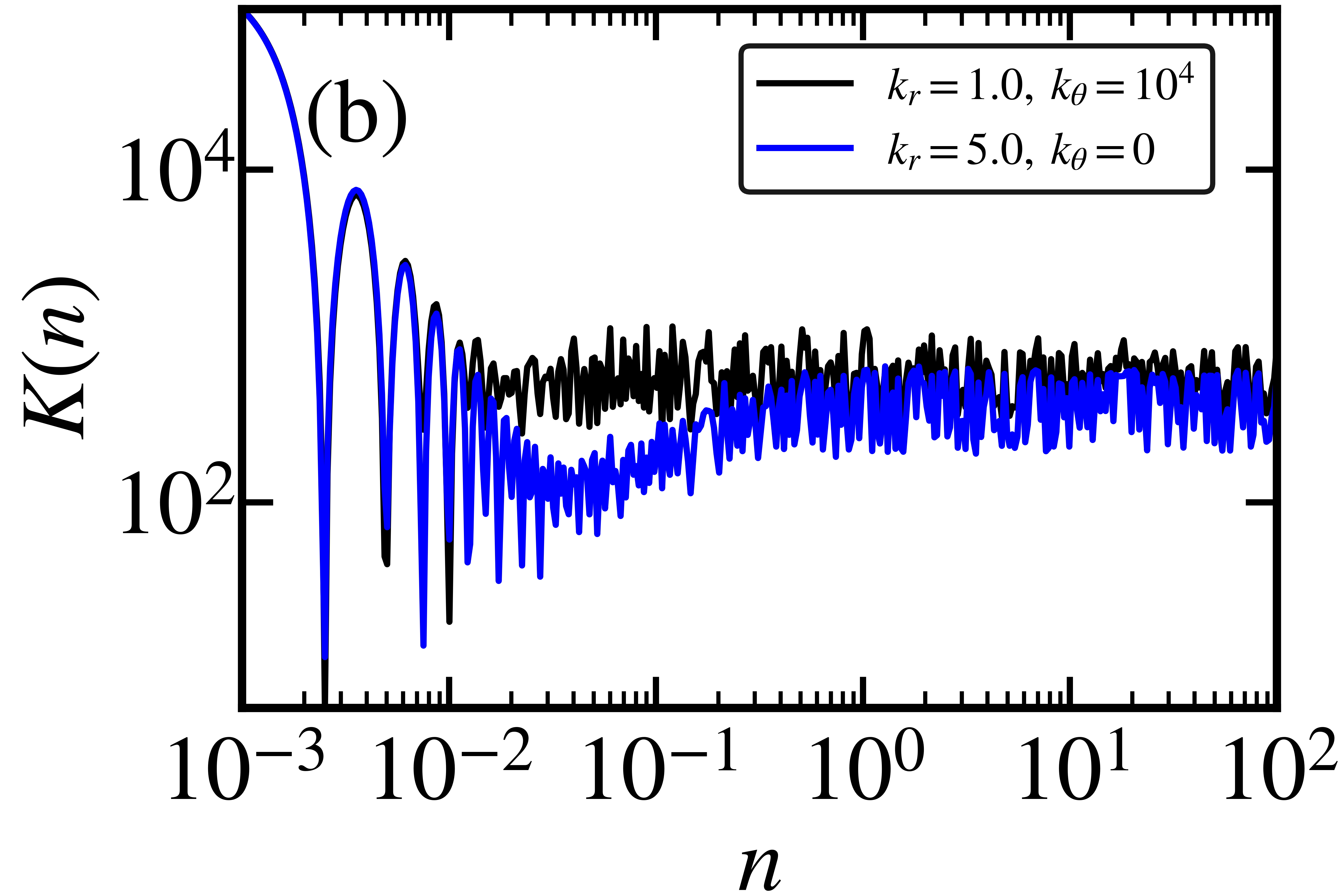}
    \caption{(a) Quantum autocorrelation function $C(n)=\frac{1}{D}\mathrm{Tr}\!\left[J_x(n)J_x(0)\right]$ (see Ref.~\cite{yashimura2025operator}) plotted as a function of time $n$ for $j=250.5$ and two representative regimes of the DKT: non-chaotic but mean-ergodic dynamics $(k_r = 1,\; k_\theta = 10^4)$ and chaotic dynamics $(k_r = 5,\; k_\theta = 0)$. Persistent oscillatory behavior in the non-chaotic case shows long-time memory and surviving correlations, whereas the rapid decay of correlations in the chaotic case signals loss of memory~\cite{nizami2024quantum,schonle2021eigenstate}. (b) SFF shows signatures of quantum chaos only for $(k_r = 5,\; k_\theta = 0)$. Here, the system size $2j=200$.}\label{supp:fig:quantum-autocorrelation}
\end{figure}
For chaotic systems with time-reversal symmetry, the SFF is expected to follow the orthogonal ensemble (GOE) prediction~\cite{Haakebook,kos2018many}. It is given by
\begin{align}\label{eq:sff-goe}
    K^{\mathrm{GOE}}(t) &= K^{\mathrm{GOE}}_{c}(t) + {\left[ \frac{\pi}{\mu t} J_1\left(\frac{2N \mu t}{\pi}\right)\right]}^2, \\[8pt]
    K^{\mathrm{GOE}}_{c}(t) &=
    \begin{cases} 
        \displaystyle
        \frac{\mu t}{\pi} - \frac{\mu t}{2\pi} \ln\!\left( 1+\frac{\mu t}{\pi} \right), & 0 < \mu t < 2\pi, \\[12pt]  
        \displaystyle 2 - \frac{\mu t}{2\pi} \ln\!\left( \frac{\mu t+\pi}{\mu t-\pi}  \right), & 2\pi < \mu t < \infty,
    \end{cases}\label{sff-goe}
\end{align}
where $\tau=t/t_H=\mu t/(2\pi)$ is the unfolded time variable, $\mu$ denotes the mean level spacing, and $t_H=2\pi/\mu$ is the Heisenberg time. The full SFF additionally contains a disconnected contribution arising from the finite spectral bandwidth, which dominates at very early times and produces the initial decay before the onset of universal correlations. The GOE ramp exhibits logarithmic corrections originating from time-reversal symmetry and is approximately twice as steep at early times (see Fig.~\ref{supp:fig:sff-ideal}). The SFF for the case $(k_r=1, k_\theta=10^4)$ decays initially and reaches a plateau without any noticeable ramp (see Fig.~\ref{supp:fig:quantum-autocorrelation}), consistent with near-integrable dynamics and the absence of long-range spectral correlations. In contrast, the chaotic case $(k_r=5, k_\theta=0)$ exhibits a clear decay $\to$ ramp $\to$ plateau structure characteristic of GOE statistics, reflecting the emergence of level repulsion and spectral rigidity associated with quantum chaotic dynamics in the presence of time-reversal symmetry.

% For chaotic systems with time-reversal symmetry, the SFF follows the Gaussian orthogonal ensemble (GOE) prediction~\cite{cotler2017black,Haakebook} given by
% \begin{align}
%     K^{\mathrm{GUE}}(t) &= K^{\mathrm{GUE}}_{c}(t) + {\left[ \frac{J_1(\mu t)}{\mu t} \right]}^2, \\[6pt]
%     K^{\mathrm{GUE}}_{c}(t) &= \begin{cases} 
%         \displaystyle \frac{\mu t}{2\pi}, & 0 < \mu t < 2\pi, \\[10pt]
%         \displaystyle 1, & 2\pi < \mu t < \infty.
%     \end{cases}\label{sff-gue}
% \end{align}
% Here, $J_1(x)$ denotes the Bessel function of the first kind, $\mu$ is the mean level spacing, and $t_H = 2\pi/\mu$ defines the Heisenberg time. The disconnected contribution, ${\left[J_1(\mu t)/(\mu t)\right]}^2$, dominates at very early times and arises from the finite spectral bandwidth, while the connected part $K_c^{\mathrm{GUE}}(t)$ gives rise to the characteristic linear ramp and plateau structure associated with quantum chaotic dynamics. The SFF for the case $(k_r=1, k_\theta=10^4)$ decays initially and reaches a plateau without any noticeable ramp (see Fig.~\ref{supp:fig:quantum-autocorrelation}). In contrast, a clear decay $\to$ ramp $\to$ plateau feature can be observed in the chaotic case $(k_r=5, k_\theta=0)$. This further supports the quantum near-integrable nature of the former case.
\end{document}